\documentclass[12pt]{article}
\pdfoutput=1
\def\Slash#1{#1\kern-0.55em\raise.05ex\hbox{/}}
\usepackage{amsmath,amssymb,amsfonts,amscd,bbm,graphicx,hyperref,longtable}
\newcommand{\Refs}{Refs.}

\newcommand{\ba}{\begin{array}}
\newcommand{\ea}{\end{array}}
\newcommand{\be}{\begin{equation}}
\newcommand{\ee}{\end{equation}}

\def\lsim{\raise0.3ex\hbox{$\;<$\kern-0.75em\raise-1.1ex\hbox{$\sim\;$}}}

\begin{document}


\thispagestyle{empty}
\renewcommand{\thefootnote}{\fnsymbol{footnote}}
\setcounter{footnote}{1}

\begin{flushright}
MAN/HEP/2010/20\\[0pt]
SISSA  41/2010/EP
\end{flushright}

\vspace*{0.5cm}

\begin{center}
\Large\bf 
Lepton Flavor Violation in Complex SUSY \\ Seesaw Models with Nearly Tribimaximal Mixing

\vspace*{15mm}

\large\bf Frank F. Deppisch$^a$\footnote{E-mail: \texttt{frank.deppisch@manchester.ac.uk}},
Florian Plentinger$^b$\footnote{E-mail: \texttt{plentinger@sissa.it}} and Gerhart
Seidl$^c$\footnote{E-mail: \texttt{seidl@physik.uni-wuerzburg.de}}
\end{center}

\vspace*{2mm}
\begin{center}
$^a${\em School of Physics and Astronomy, University of Manchester}\\
{\em Manchester M13 9PL, United Kingdom}
~\\\vspace*{3mm}
$^b${\em SISSA and INFN-Sezione di Trieste}\\
{\em via Bonomea 265, 34136 Trieste, Italy}
~\\\vspace*{3mm}
$^c${\em Institut f\"ur Theoretische Physik und Astrophysik}\\
{\em Universit\"at W\"urzburg, 97074 W\"urzburg, Germany} 
\end{center}

\vspace*{15mm}

\centerline{\bf Abstract}
\vspace*{2mm}
We survey the lepton flavor violation branching ratios $\text{Br}(\mu\to e\gamma)$, $\text{Br}(\tau\to \mu\gamma)$, and
$\text{Br}(\tau\to e\gamma)$ in mSUGRA for a broad class of lepton mass matrix textures that give nearly tribimaximal lepton mixing. Small neutrino masses are generated by the type-I seesaw mechanism with non-degenerate right-handed neutrino masses. The textures exhibit a hierarchical mass pattern and can be understood from flavor models giving rise to large leptonic mixing. We study the branching ratios for the most general CP-violating forms of the textures. It is demonstrated that the branching ratios can be enhanced by 2-3 orders of magnitude as compared to the CP-conserving case. The branching ratios exhibit, however, a strong dependence on the choice of the phases in the Lagrangian which affects the significance of flavor models. In particular, for general CP-phases, the lepton flavor violating rates appear to be essentially uncorrelated with the possible high- and low-energy lepton mixing parameters, such as the reactor angle.

\renewcommand{\thefootnote}{\arabic{footnote}}
\setcounter{footnote}{0}

\newpage


\section{Introduction}
\label{sec:introduction}
Neutrino oscillation experiments have during the past decade pinned down the neutrino mass and mixing parameters to a remarkable precision \cite{Fukuda:2002pe,Fukuda:1998mi,Araki:2004mb,Aliu:2004sq}. Global fits \cite{Schwetz:2008er} tell us that the solar and atmospheric neutrino mass squared differences are (at $1\sigma$)
\begin{equation}\label{eq:masssquareddiffs}
\Delta m_\odot^2=7.65^{+0.23}_{-0.20}\times 10^{-5}\,\text{eV}^2,\quad
|\Delta m_\text{atm}^2|=2.40^{+0.12}_{-0.11}\times 10^{-3}\,\text{eV}^2,
\end{equation}
whereas the solar, atmospheric and reactor\footnote{Note, the current best-fit value for the reactor angle is $\theta_{13}=6.5^\circ$.} mixing angle are
respectively given by
\begin{equation}\label{eq:mixingangles}
\theta_{12}=(32.6^{+3.2}_{-2.7})^\circ,\quad
\theta_{23}=(45.0^{+4.1}_{-10.3})^\circ,\quad
\theta_{13} \leq 13.2^\circ\,\,(3\sigma).
\end{equation}
The leptonic Pontecorvo-Maki-Nakagawa-Sakata (PMNS) mixing matrix $U_\text{PMNS}$ \cite{PMNS} can be well
approximated by the Harrison-Perkins-Scott (HPS) tribimaximal mixing matrix $U_\text{HPS}$
\cite{Harrison:1999cf} (up to phases) as
\begin{equation}\label{eqU:HPS}
U_\text{PMNS}\approx U_\text{HPS}=
\left(
\begin{matrix}
\sqrt{\frac{2}{3}} & \frac{1}{\sqrt{3}} & 0\\
-\frac{1}{\sqrt{6}} & \frac{1}{\sqrt{3}} & \frac{1}{\sqrt{2}}\\
\frac{1}{\sqrt{6}} & -\frac{1}{\sqrt{3}} & \frac{1}{\sqrt{2}}
\end{matrix}
\right).
\end{equation}
In $U_\text{HPS}$, the solar and the atmospheric angle are given by $\theta_{12}\approx 35^\circ$ and
$\theta_{23}=45^\circ$, whereas the reactor angle $\theta_{13}$ vanishes. The measured PMNS mixing
angles may thus be treated as deviations from exact tribimaximal mixing \cite{Plentinger:2005kx,Majumdar:2006px}, describing nearly tribimaximal lepton mixing \cite{Xing:2002sw}.

The smallness of the absolute neutrino mass scale $m_\nu\simeq 5\times 10^{-2}\,\text{eV}$ can be understood in terms of seesaw mechanism \cite{typeIseesaw,typeIIseesaw}, which establishes a connection between low-energy
neutrino observables and physics near the unification scale $M_X\approx 2.5\times 10^{16}\,\text{GeV}$ \cite{GUTscale}. In such a context, neutrino oscillations can therefore probe high-scale theories of flavor.

It is interesting to ask whether there are possibilities to test mass and lepton flavor models in other than only neutrino oscillation experiments. Such a possibility is offered by charged lepton flavor violation (LFV) in supersymmetry (SUSY). In the standard model (SM), LFV is absent and adding right-handed neutrinos to the SM only tiny LFV effects that are suppressed by the smallness of the neutrino masses. In SUSY, however, virtual effects of the right-handed neutrinos and their superpartners affect the renormalization group equations (RGE) of the slepton masses and trilinear couplings and induce observable branching ratios (BRs) of LFV decays \cite{Borzumati:1986qx}.

One attractive feature of the seesaw mechanism is that it allows for leptogenesis via the decay of the right-handed neutrinos \cite{Fukugita:1986hr} (for reviews see \cite{leptogenesis}). Barring special cases, however, the complex Yukawa couplings providing the necessary CP-violating phases can generally lead to a large enhancement of the LFV rates by large factors $\gtrsim 10^4$ \cite{Albino:2006xe}. If these CP phases are not fixed, their presence may obscure possible predictions from flavor models and, furthermore, make it difficult to distinguish experimentally different candidate theories of flavor. For a discussion of LFV in certain quark-lepton complementarity scenarios see for example \cite{Hochmuth:2006xn}.

In this paper, we study the LFV decays $\mu\rightarrow e\gamma$, $\tau\to \mu\gamma$, and $\tau\to e\gamma$,  in minimal supergravity (mSUGRA) for a large class of CP-conserving and CP-violating lepton mass matrix textures that yield nearly tribimaximal lepton mixing. The textures are formulated at the level of the Lagrangian and are characterized by hierarchical entries that can be generated in flavor models using, e.g.~the Froggatt-Nielsen (FN) mechanism \cite{Froggatt:1978nt}. The textures implement the idea of quark-lepton complementarity (QLC) \cite{qlc} in a general way: They realize nearly tribimaximal lepton mixing by taking large contributions from different lepton sectors, i.e.~the charged lepton and the left- and right-handed neutrino sector, into account. The FN mechanism has, for instance, been used in \Refs~\cite{Plentinger:2006nb,Plentinger:2007px,Plentinger:2008up,Winter:2007yi} to construct charged lepton and even neutrino mass textures which can be implemented by discrete flavor symmetries~\cite{Plentinger:2008up,Plentinger:2008nv,Altarelli:2008bg}. Many models have been proposed in the literature to reproduce tribimaximal leptonic mixing using non-Abelian flavor symmetries (for early models based on $A_4\simeq Z_3\ltimes(Z_2\times Z_2)$ and examples using the double covering group of $A_4$, see Refs.~\cite{A4} and \cite{2A4}).
We demonstrate that the inclusion of random CP-violating phases at the level of the Lagrangian can enhance the branching ratios typically up to 2-3 orders of magnitude. Moreover, we find a strong dependence of the LFV-rates on
the choice of the phases in the Lagrangian. As a consequence, it appears that the assignment of random CP-violating
phases in the Lagrangian in general erase possible correlations between LFV rates and the PMNS mixing parameters of the low scale theory.

The paper is organized as follows: In Sec.~\ref{sec:seesaw}, we introduce the SUSY seesaw mechanism and its different parametrizations. Charged lepton flavor violation in SUSY is described in Sec.~\ref{sec:lfv}. The types of textures for which we study LFV are introduced in Sec.~\ref{sec:textures}, and our results for LFV are then presented in
Sec.~\ref{sec:results}. Finally, in Sec.~\ref{sec:summaryandconclusions}, we give our summary and conclusions.

\section{SUSY Seesaw Mechanism}\label{sec:seesaw} 

In what follows, we assume that the left-handed active neutrinos acquire their masses via the type-I seesaw mechanism
\cite{typeIseesaw}. We consider the seesaw mechanism in SUSY, which differs from the usual seesaw mechanism in the respect that it involves two Higgs doublets. In the type-I seesaw mechanism, the part of the superpotential generating the charged lepton and neutrino masses is given by
\begin{equation}\label{eq:superpotential}
 W=-(Y_\ell)_{ij}e^c_i\ell_j H_1-(Y_D)_{ij}\nu^c_i\ell_jH_2+\frac{1}{2}(M_R)_{ij}\nu^c_i\nu^c_j,
\end{equation}
where $\ell_i=(\nu_i,\:e_i)^T$, $e_i^c$, and $\nu^c_i$ ($i=1,2,3$
is the generation index) are the matter superfields of the left-handed lepton doublets, right-handed charged
leptons, and right-handed SM singlet neutrinos, respectively. In (\ref{eq:superpotential}), $H_1$ and $H_2$ are the
usual Higgs superfield doublets generating the down- and up-type masses, respectively, and $Y_\ell$ and $Y_D$ are
the $3\times 3$ Dirac Yukawa coupling matrices of the charged leptons and neutrinos, whereas $M_R$ is the $3\times 3$
Majorana mass matrix of the right-handed neutrinos. After electroweak symmetry breaking, the Higgs doublets develop
vacuum expectation values $\langle H^0_i\rangle$, where $\langle H_2^0 \rangle = v\sin\beta$ with $v=174\,\text{GeV}$ and $\tan\beta =\langle H_2^0\rangle/\langle H_1^0\rangle$. The resulting lepton mass terms become
\begin{equation}\label{eq:massterms}
\mathcal{L}_\text{mass}=-(M_\ell)_{ij}e^c_ie_j
 -(M_D)_{ij}\nu^c_i\nu_j+\frac{1}{2}(M_R)_{ij}\nu^c_i\nu^c_j+\text{h.c.},
\end{equation}
where $M_\ell=\langle H_1^0\rangle Y_\ell$ is the charged lepton and $M_D=\langle H_2^0\rangle Y_D$ the Dirac neutrino mass matrix. $M_\ell$ and $M_D$ are complex $3\times 3$ matrices. The matrix $M_R$ is complex and symmetric and has matrix elements of the order of the $B-L$ breaking scale $M_{B-L}\sim 10^{14}\:\text{GeV}$. After integrating out the heavy right-handed neutrinos, we obtain the effective low-energy $3\times 3$ neutrino Majorana mass matrix
\begin{equation}
\label{eq:Meff}
 	M_\text{eff}=M_D^TM_R^{-1}M_D=Y_D^T M^{-1}_R Y_D (v \sin\beta )^2,
\end{equation}
giving masses $\sim 10^{-2}\;\text{eV}$ to the light neutrinos in agreement with observation. The seesaw mechanism is attractive since it establishes a connection between the absolute neutrino mass scale and $M_{B-L}$, which is close to the grand unified theory (GUT) scale $M_X\sim 10^{16}\,\text{GeV}$.

The mass terms in (\ref{eq:massterms},\ref{eq:Meff}) are diagonalized by unitary 3$\times$3 matrices $U_x$ 
which in their most general form can be written as (cf.~\Refs~\cite{Plentinger:2006nb,Plentinger:2007px})
\begin{subequations}
\begin{equation}
 U_x= D_x \cdot \widehat{U}_x \cdot K_x,\quad\quad x=\ell,\ell',D,D',R,\nu,
\label{eq:DUK}
\end{equation}
where $D_x$ and $K_x$ are given by $D_x=\text{diag}(e^{\text{i}\varphi_1},e^{\text{i}\varphi_2},e^{\text{i}\varphi_3})$
and $K_x=\text{diag}(e^{\text{i}\alpha_1},e^{\text{i}\alpha_2},1)$ with $\varphi_i\in[0,2\pi[$ and $\alpha_i\in[0,\pi[$ and
\begin{equation}
 \label{eq:ckm}
 \widehat{U}_x = \left(
 \begin{array}{ccc}
   c_{12} c_{13} & s_{12} c_{13} & s_{13} e^{-\text{i}\widehat{\delta}} \\
   -s_{12} c_{23} - c_{12} s_{23} s_{13} e^{\text{i}\widehat{\delta}} &   c_{12} c_{23} -
 s_{12} s_{23} s_{13} e^{\text{i}\widehat{\delta}} & s_{23} c_{13} \\
 s_{12} s_{23} - c_{12} c_{23} s_{13}
e^{\text{i}\widehat{\delta}} & -c_{12} s_{23} - s_{12} c_{23} s_{13} e^{\text{i}\widehat{\delta}} & c_{23}
c_{13}
 \end{array}
 \right)
\end{equation}
\end{subequations}
is a unitary matrix in the standard parametrization with $s_{ij} =
 \sin\hat{\theta}_{ij}$, $c_{ij} = \cos\hat{\theta}_{ij}$, where
$\hat{\theta}_{ij}\in\{\hat{\theta}_{12},\hat{\theta}_{13},\hat{\theta}_{23}\}\in\left[0,\frac{\pi}{2}\right]$, and
$\widehat{\delta}\in[0,2\pi[$. The mass matrices then read
\begin{equation}\label{eq:massmatrices}
M_\ell=U_{\ell'}^\ast M_\ell^\text{diag}U_{\ell}^T,\quad
M_D = U_{D'}^\ast M_D^\text{diag} U_{D}^T,\quad M_R = U_R
M_R^\text{diag} U_R^T,\quad M_\text{eff}=U_\nu M_\text{eff}^\text{diag}U_\nu^T,
\end{equation}
where the diagonal mass matrices
\begin{subequations}
\begin{eqnarray}
M_\ell^\text{diag}=\text{diag}(m_e,m_\mu,m_\tau),&&
M_D^\text{diag}=\text{diag}(m_1^D,m_2^D,m_3^D),\\
M_R^\text{diag}=\text{diag}(m_1^R,m_2^R,m_3^R),&&
M_\text{eff}^\text{diag}=\text{diag}(m_1,m_2,m_3),
\end{eqnarray}
\end{subequations}
having positive mass eigenvalues. The PMNS mixing matrix is given by
\begin{equation}\label{eq:PMNS}
U_\text{PMNS}=U_\ell^\dagger U_\nu=\widehat{U}_\text{PMNS}K_\text{Maj},
\end{equation}
where the matrix $\widehat{U}_\text{PMNS}$ in (\ref{eq:PMNS}) is described in the
 standard parametrization (\ref{eq:ckm}) by the solar angle $\hat\theta_{12}=\theta_{12}$, the reactor angle
$\hat\theta_{13}=\theta_{13}$, the atmospheric angle $\hat\theta_{23}=\theta_{23}$ and the Dirac CP-phase $\hat\delta=\delta$. $K_\text{Maj}=\text{diag}(e^{\text{i}\phi_1},e^{\text{i}\phi_2},1)$ contains the Majorana phases $\phi_{1}$ and $\phi_2$ with $\phi_{1,2}\in[0,\pi[$.\footnote{The phase matrix $K_\ell$ has been absorbed into the right-handed charged lepton sector.}

Rotating the lepton doublets $\ell_i$ by $U_\ell$ to the basis where $M_\ell$ is diagonal, $M_{\text{eff}}$ is diagonalized by the PMNS matrix
\begin{equation}\label{eq:NeutrinoDiag}
U_\text{PMNS}^\dagger M_\text{eff} U_\text{PMNS}^\ast= M_\text{eff}^\text{diag}.
\end{equation}
Unless stated otherwise, we will from now on work in the basis where the charged lepton and heavy Majorana neutrino mass matrix are diagonal, i.e. where $M_\ell=M_\ell^\text{diag}$ and $M_R=M_R^\text{diag}$.

The significance of the mixing matrices $U_x$ in (\ref{eq:massmatrices}) is that the mass terms in (\ref{eq:massterms}) may be predicted by some flavor model such as the Froggatt-Nielsen mechanism \cite{Froggatt:1978nt} or discrete flavor symmetries~\cite{Plentinger:2008up,Plentinger:2008nv,Altarelli:2008bg}. Therefore, by tracing $U_\text{PMNS}$ back to the matrices $U_x$, which describe the
high-energy lepton mixing, one can gain better understanding of the observable PMNS parameters in terms of a possible
fundamental theory of flavor.

Flavor models do not fix the absolute scale of the Yukawa coupling matrix. In order to properly normalize all couplings, we shall express $Y_D$ as a function of the absolute neutrino mass scale $m_\nu\sim 10^{-2}\;\text{eV}$ and $M_{B-L}\sim 10^{14}\,\text{GeV}$. For this purpose, consider arbitrary $M_\text{eff}$ and $M_R$ and let $m_3$ and $m^R_3$ be the heaviest mass eigenvalues of $M_\text{eff}$ and $M_R$, respectively. We normalize $M_\text{eff}$ by redefining
\begin{equation}\label{eq:normalization}
M_\text{eff}\rightarrow M_\text{eff}'=
\frac{m_\nu}{m_3}M_\text{eff}.
\end{equation}
Adjusted in this way, the heaviest mass eigenvalue of $M_\text{eff}$ becomes equal to $m_\nu$. Similarly, we re-scale 
$m_3^R=M_{B-L}$ to normalize $M_R$. The normalization of $M_\text{eff}$ in (\ref{eq:normalization}) is then equivalent to the redefinitions
\begin{equation}\label{eq:redefinitions}
 M_R\rightarrow M_R'=\frac{M_{B-L}}{m^R_3}M_R\quad\text{and}\quad
 M_D\rightarrow
 M_D'=\sqrt{\frac{m_\nu}{m_3}\frac{M_{B-L}}{m_3^R}}M_D=\langle
 H^0_2\rangle Y_D,
\end{equation}
where $Y_D$ is now properly normalized. From (\ref{eq:massmatrices}) and
(\ref{eq:DUK}) we find that in this basis
\begin{equation}\label{eq:MD}
M_D=K_R^\ast\widehat{U}_R^\dagger
\tilde{D}\widehat{U}_{D'}^*
M_D^{\mathrm{diag}}\tilde{K}\widehat{U}_D^T D_DU_\ell^\ast.
\end{equation}
For simplicity, we will make this rescaling in the following without noting the prime in $M_R$ and $M_D$. From
(\ref{eq:MD}) 
and (\ref{eq:redefinitions}), 
we see that
\begin{equation}\label{eq:YD}
 Y_D=
\frac{1}{\langle H_2^0\rangle}
\sqrt{\frac{m_\nu}{m^3_\nu}\frac{M_{B-L}}{m_3^R}}
K^\ast_R\widehat{U}_R^\dagger \tilde{D}\widehat{U}_{D'}^*
M_D^{\mathrm{diag}}\tilde{K}\widehat{U}_D^T D_DU_\ell^\ast,
\end{equation}
where we have introduced $\tilde{K}=K_DK_{D'}^\ast$ and $\tilde{D}=D_{D'}^\ast D_R^\ast$.
The neutrino Yukawa coupling matrix in (\ref{eq:YD}) can also be written as follows \cite{Casas:2001sr}\footnote{Note that the   definition of $U_\text{PMNS}$ in \cite{Casas:2001sr} differs from our definition by complex conjugation (cf.~(\ref{eq:NeutrinoDiag})).}:
\begin{equation}\label{eq:CI}
    Y_D =
    \frac{1}{v\sin\beta}\sqrt{M_R^\text{diag}}\cdot R
\cdot \sqrt{M_\text{eff}^\text{diag}}\cdot U^T_\text{PMNS}.
\end{equation}
Here, $R$ denotes a complex orthogonal matrix  which may be
parametrized in terms of 3 complex angles $\theta_i=x_i +\text{i}y_i$ as
\begin{equation}\label{eq:RMatrix}
R=\left(\begin{array}{ccc}
c_2 c_3 & -c_1 s_3-s_1 s_2 c_3 & s_1 s_3 -c_1 s_2 c_3\\
c_2 s_3 & c_1 c_3 -s_1 s_2 s_3 & -s_1 c_3 -c_1 s_2 s_3 \\
s_2 & s_1 c_2 & c_1 c_2 \end{array}\right),
\end{equation}
with $(c_i, s_i)=(\cos \theta_i,\sin \theta_i) =(\cos x_i \cosh
y_i -\text{i}\sin x_i \sinh y_i, \sin x_i \cosh y_i +\text{i}\cos x_i \sinh
y_i)$. The parameters can take the values $x_i\in [0,2\pi[$ and $y_i\in \, {] -\infty,\infty[}$ (in practical cases, however, the $y_i$ are constrained by perturbativity to values $|y_i|\lesssim\mathcal{O}(1)$, see Sec.~\ref{sec:CPviolating}). While the light neutrino masses $m_i$ and the mixing angles $\theta_{ij}$ have been measured or constrained, the phases $\phi_i$ and $\delta$, the heavy neutrino masses $m_i^R$ and the matrix $R$ are presently unknown. From (\ref{eq:YD}) and (\ref{eq:CI}) we obtain
\begin{equation}\label{eq:connection}
 R=
(M_R^\text{diag})^{-1/2}
K^\ast_R\widehat{U}_R^\dagger \tilde{D}\widehat{U}_{D'}^*
M_D^{\mathrm{diag}}\tilde{K}\widehat{U}_D^T D_D\widehat{U}_\nu^\ast K_\text{Maj}^\ast(M_\text{eff}^\text{diag})^{-1/2},
\end{equation}
where we have used (\ref{eq:PMNS}).
The parametrization in (\ref{eq:CI}) has the advantage that (i) one can understand the
impact of nonzero CP-violating phases on the LFV rates in a comparatively simple way and (ii)
one can quickly scan the parameter space while ensuring a valid low-energy phenomenology (lepton masses and PMNS
angles). We will make use of these properties in subsequent sections. Note, however, that the drawback of the parametrization (\ref{eq:CI}) is that the exact connection with the lepton mass terms in the Lagrangian has been lost after rotating to the basis where $M_\ell$ and $M_R$ are diagonal.

\section{Charged Lepton Flavor Violation in SUSY}\label{sec:lfv}
The heavy neutrino mass eigenstates $\nu^c_i$ introduced in the seesaw mechanism  are too heavy to be observed directly but they influence the mixing of the sleptons in the MSSM via radiative corrections. The $6\times 6$ slepton mass matrix may be written as a sum of two parts,
\begin{eqnarray}\label{eq:sleptonmassmatrix}
 m_{\tilde l}^2=\left(
	\begin{array}{cc}
		m_L^2    & m_{LR}^{2\dagger} \\
		m_{LR}^2 & m_R^2
	\end{array}
	\right)_{\rm MSSM}+\left(
	\begin{array}{cc}
		\delta m_L^2    & \delta m_{LR}^{2^\dagger} \\
		\delta m_{LR}^2 & 0
	\end{array}
	\right)_{\nu^c},
\end{eqnarray}
where the first part denoted by MSSM is the usual mass matrix in the MSSM without right-handed neutrinos. In our analysis of SUSY LFV processes we adopt the mSUGRA scheme of SUSY-breaking, in which case the slepton mass matrix does not contain flavor mixing terms. In leading logarithmic approximation the corrections to the slepton mass
matrix due to right-handed neutrinos, denoted by $\nu^c$ in (\ref{eq:sleptonmassmatrix}), can then be written as \cite{Hisano:1999fj},
\begin{eqnarray}
	\label{left_handed_SSB2}
	\delta m_{L}^2  &=& 
	-\frac{1}{8 \pi^2}(3m_0^2+A_0^2)Y_D^\dag L Y_D, \nonumber\\
	\delta m_{LR}^2 &=&
	 -\frac{3}{16\pi^2} A_0 v \cos\beta Y_l Y_D^\dag L Y_D,
\end{eqnarray}
where $L_{ij} = \ln(M_X/m_i^R)\delta_{ij}$, $m^R_i$ being the heavy neutrino masses, and \(m_0\) and \(A_0\) are the universal scalar mass and trilinear coupling, respectively, at $M_X$. With the neutrino Yukawa matrix $Y_D$ and the heavy neutrino masses $m^R_i$ in a given scenario as input at the GUT scale we calculate the slepton mass matrix at the electroweak scale using (\ref{left_handed_SSB2}). The flavor off-diagonal virtual effects in (\ref{left_handed_SSB2}) induced by the mixing in the neutrino sector lead to charged LFV. More details on this mechanism can be found in \cite{Deppisch:2002vz} and the references therein.

\subsection{LFV Rare Decays}\label{sec:RareDecays}
In the SUSY seesaw model considered here, LFV processes mainly occur via intermediate left-handed slepton flavor transitions. The most important low-energy processes are the rare decays $l_i\to l_j\gamma, i \neq j \in e,\mu,\tau$, which provide the most stringent bounds on LFV in the SUSY seesaw model as of now. The current bounds on these processes as well as the expected sensitivities of future experiments are listed in Table~\ref{Br23fromBr12seesawforpresentfuture}. 

Each LFV transition is suppressed in a given process by a small factor $|(\delta m_L^2)_{ij}/\widetilde{m}^2|^2$ ($i\neq j$), where $(\delta m_L^2)_{ij}$ are the off-diagonal elements of the left-handed slepton mass matrix $m_L^2$ specified in (\ref{left_handed_SSB2}) and $\widetilde{m}^2$ is of the order of the relevant sparticle masses in the loops involved in the process. To lowest order in the off-diagonal mass corrections one has
approximately~\cite{Hisano:1999fj},
\begin{equation}\label{eqn:DecayApproximation}
	\text{Br}(l_i \to l_j \gamma) \propto \alpha^3 m_{l_i}^5
	\frac{|(\delta m_L^2)_{ij}|^2}{\widetilde{m}^8}\tan^2\beta.
\end{equation}
This expression is just used for illustration. In our numerical calculations we use the full one loop result for \(\text{Br}(l_i \to l_j \gamma)\), as given in \cite{Deppisch:2002vz}.

\begin{table}[t]
\begin{center}
\begin{tabular}{|c|c|c|c|}
\hline 
 & Br$(\mu\to e\gamma)$ & Br$(\tau\to\mu\gamma)$ & Br$(\tau\to e\gamma)$ \\
\hline
Present & $1.2 \times 10^{-11}$\cite{Eidelman:2004wt} & $6.8 \times 10^{-8}$ \cite{Aubert:2005ye} & $1.1 \times 10^{-7}$ \cite{Aubert:2005wa}\\
\hline
Expected & $10^{-13}$\cite{Nicolo:2009zz} & $\approx 10^{-8}$ & $\approx 10^{-8}$ \\
\hline
\end{tabular}
\end{center}
\caption{Current bounds and expected future sensitivities of direct experimental LFV searches.} \label{Br23fromBr12seesawforpresentfuture}
\end{table}

\subsection{LFV Processes at the LHC}\label{sec:lfvatlhc}
At the LHC, a feasible test of LFV is provided by the production of squarks and gluinos, followed by cascade decays of squarks and gluinos via neutralinos and sleptons
\cite{Agashe:1999bm,Andreev:2006sd}:
\begin{eqnarray}\label{eqn:LHCProcesses}
    pp             &\to& \tilde q_a \tilde q_b, \tilde g \tilde q_a, \tilde g
    \tilde g,\nonumber\\
    \tilde q_a(\tilde g)&\to& \tilde\chi^0_2 q_a(g),\nonumber \\
    \tilde\chi^0_2 &\to& \tilde l_c l_i,\nonumber\\
    \tilde l_c     &\to& \tilde\chi^0_1 l_j,
\end{eqnarray}
where \(a,b,c\) run over all relevant sparticle mass eigenstates. Lepton flavor violation can occur in the decay of the second lightest neutralino or the slepton, resulting in different lepton flavors, \(i\neq j\). The total cross section for the signature \(l^\pm_i l^\mp_j + X\) can then be written as
\begin{eqnarray}
\label{eqn:LHCProcessA}
       \sigma&&\!\!\!\!\!\!\!\!\!\!\!\!\!(pp\to \tilde\chi_2^0 + X \to l_i^\pm l_j^\mp\tilde\chi_1^0 + X)  \nonumber\\
	&=&\!\!\!\left\{\phantom{\sum_a}\!\!\!\!\!\!\!\!2
                  \sigma(pp \to \tilde g\tilde g)
                  \text{Br}(\tilde g\to qq\tilde\chi_2^0)\right. \nonumber\\
        &+&\!\!\!\sum_a \sigma(pp\to\tilde g\tilde q_a) 
      \left[
                   \text{Br}(\tilde g  \to qq\tilde\chi_2^0)
                 + \text{Br}(\tilde q_a\to q\tilde\chi_2^0)
           \right]                                          \nonumber\\
        &+&\!\!\!\left.\sum_{a,b}\sigma(pp \to \tilde q_a\tilde q_b)
           \left[
                   \text{Br}(\tilde q_a\to q\tilde\chi_2^0)
                 + \text{Br}(\tilde q_b\to q\tilde\chi_2^0)
           \right]\right\}
	\times Br(\tilde\chi_2^0\to l_i^\pm l_j^\mp\tilde\chi_1^0),
\end{eqnarray}
with $\text{Br}(\tilde g\to qq\tilde\chi_2^0)=\sum_a Br(\tilde g\to q\tilde q_a)Br(\tilde q_a\to q\tilde\chi_2^0)$  and
 \(X\) can involve jets, leptons and lightest neutralinos produced by lepton flavor conserving decays of squarks and gluinos, as well as low-energy proton remnants. The LFV branching ratio \(\text{Br}(\tilde\chi^0_2\to
l_i^+l_j^-\tilde\chi^0_1)\) is for example calculated in \cite{Bartl:2005yy} in the framework of model-independent MSSM slepton mixing. In general, it involves a coherent summation over all intermediate slepton states. In our numerical calculation we use the leading order partonic cross sections $\sigma(pp\to\tilde q\tilde q)$, $\sigma(pp\to\tilde q\tilde g)$ and $\sigma(pp\to\tilde g\tilde g)$ folded with the CTEQ6M parton
distribution functions together with (\ref{eqn:LHCProcessA}) in order to calculate the number of events for the LFV process $pp\to \tilde\chi_2^0 + X \to e\mu + \tilde \chi_1^0 + X$ expected at the LHC \cite{Deppisch:2007rm,Raidal:2008jk}. 

\section{Textures}\label{sec:textures}
In the following, we will consider the lepton mass matrix textures from the list of 72 types given in
\cite{Plentinger:2007px}, where we normalize the textures according to (\ref{eq:redefinitions}) to obtain lepton and neutrino masses compatible with experimental data. This list will be called our ``reference list'' of textures. Each of the 72 types of textures in the list is characterized by a set $\{M_\ell,M_D,M_R\}$ of relevant mass matrices. We call this set of three textures a ``texture set''. It is important to note that the textures $M_\ell,M_D,$ and $M_R$ are in general not diagonal and large leptonic mixing can emerge from any of the matrices $M_\ell,M_D$ or $M_R$. The structure of the
non-diagonal textures is relevant for the construction of explicit high-scale theories, such as the Froggatt-Nielsen mechanism or discrete flavor models, generating the hierarchical pattern of these textures. In fact, a large number of explicit models predicting the texture sets from flavor symmetries have already been found for the SM \cite{Plentinger:2008up} and also for SUSY $SU(5)$ GUTs \cite{Plentinger:2008nv}.

\subsection{Real Textures and Relation to QLC}\label{sec:CPconserving}
As shown in \cite{Plentinger:2007px}, each of the 72 texture sets reproduces tribimaximal neutrino mixing in the neutrino sector at the $3\sigma$ level along with the charged lepton mass ratios\footnote{We are interested in a fit compatible with minimal $SU(5)$ but different mass spectra, e.g. realization of the Georgi-Jarlskog relations \cite{Georgi:1979df} are possible and can be implemented just as well.}
\begin{equation}\label{eq:chargedleptonmasses}
m_e:m_\mu:m_\tau=\epsilon^4:\epsilon^2:1,
\end{equation}
and a normal neutrino mass hierarchy of the form
\begin{equation}\label{eq:neutrinomasses}
m_1:m_2:m_3=\epsilon^2:\epsilon:1,
\end{equation}
where $\epsilon$ is of the order of the Cabibbo angle $\epsilon\simeq \theta_\text{C}\simeq 0.2$ and $m_3=m_\nu\approx 5\times 10^{-2}\,\text{eV}$, which reproduces the values of the neutrino mass
squared differences in (\ref{eq:masssquareddiffs}). In the reference list, $M_D$ has the eigenvalues $m_i^D$ and $M_R$ the eigenvalues $m_i^R$ ($i=1,2,3$). Their ratios are always of the form
\begin{equation}\label{eq:leptonmasses}
 m_1^D:m_2^D:m_3^D=\epsilon^k:\epsilon^m:\epsilon^n, \quad
 m_1^R:m_2^R:m_3^R=\epsilon^p:\epsilon^q:1,
\end{equation}
where $k,m,n,p,q$ are non-negative integers $\leq 2$ and $0< p\leq q$. The heavy Majorana neutrino masses are, thus, always non-degenerate. The textures in the reference list lead to solar and atmospheric mixing angles that are in agreement with current data at the $3\sigma$ level. Additionally, the reactor angle is very small and satisfies $\theta_{13}<1^\circ$ which is an attractive parameter range for flavor models although new global fits point slightly towards a nonzero $\theta_{13}$.

The textures give rise to a large solar angle $\theta_{12}\approx 33^\circ$ in a way similar to QLC. This means that the mixing angles entering the mixing matrices $U_x$ in (\ref{eq:massmatrices}) are either of the order of $\sim\epsilon^n$, with a positive integer $n$, or they are equal to $\pi/4$, corresponding to maximal mixing. The nearly tribimaximal mixing form of $U_\text{PMNS}$, and in particular the observed value of the solar angle, is then a consequence of combining the mixing matrices $U_x$ from different lepton sectors, such as the charged lepton or left-handed neutrino sector, into $U_\text{PMNS}$ via (\ref{eq:PMNS}).

Lepton mixing angles with positive integer powers of $\epsilon$ are motivated by the observed CKM mixing
$V_{us}\simeq\epsilon,V_{cb}\simeq\epsilon^2,$ and $V_{ub}\simeq\epsilon^3$, and by $\mu$-$\tau$-symmetry
\cite{earlymutau} (for more recent studies on $\mu$-$\tau$-symmetry see, e.g. \cite{recentmutau}). Differently from most
applications of QLC, however, we do not require that $U_\ell$ be of a CKM-like mixing form (i.e.~that $U_\ell\simeq V_\text{CKM}$) or that $U_\nu$ be of the bi-maximal mixing type. Instead, in our reference list, the observed large solar and atmospheric mixing angles in $U_\text{PMNS}$ can originate from maximal mixing among any two generations of charged leptons, left-handed, or right-handed neutrinos. For instance, we have in 10\% of the cases {\it trimaximal} mixing of the left-handed charged leptons, i.e.~all three mixing angles in $U_\ell$ are maximal. Such general forms of textures as given in the reference list are useful for the construction of new explicit models that can explain the observed lepton mass and mixing parameters.

Consider, for example, texture set No.~1 from the reference list \cite{Plentinger:2007px}. The charged lepton
texture reads
\begin{equation}\label{eq:Ml}
 M_\ell=\frac{m_\tau}{\sqrt{2}}
\left(
\begin{matrix}
-a\epsilon^4 & 0 & 0\\
a\epsilon^4 & \epsilon^2 & -\epsilon^2\\
a\epsilon & 1 & 1 
\end{matrix}
\right),
\end{equation}
while the neutrino mass matrix textures are
\begin{equation}\label{eq:MDMR}
M_D=-\frac{m_3^D}{\sqrt{2}}
\left(
\begin{matrix}
\epsilon^2 & a\epsilon^2 & -\epsilon^2\\
a\epsilon & 1 & \epsilon^2\\
a\epsilon^4 & -1 & a\epsilon
\end{matrix}
\right), \quad
M_R=\frac{m_3^R}{2}\left(
\begin{matrix}
b\epsilon^2 & a\epsilon^3 & a\epsilon^3\\
a\epsilon^3 & 1+\epsilon & -1+\epsilon\\
a\epsilon^3 & -1+\epsilon & 1+\epsilon
\end{matrix}
\right),
\end{equation}
where the order-one coefficients $a$ and $b$, take the values $a=\sqrt{2},$ and $b=2$. The mass parameter $m_3^D\sim 10^2\,\text{GeV}$ is determined by a normalization as described in (\ref{eq:redefinitions}). For this example, the ratios of the corresponding eigenvalues are
\begin{equation}
m_1^D:m_2^D:m_3^D=\epsilon^2:1:\epsilon,\quad 
m_1^R:m_2^R:m_3^R=\epsilon^2:\epsilon:1.
\end{equation}
The charged lepton and light neutrino mass ratios are as in (\ref{eq:chargedleptonmasses}) and (\ref{eq:neutrinomasses}). This texture set is an example where we have maximal mixing among the 2nd and 3rd generation of left-handed charged leptons, $\theta_{23}^\ell=\pi/4$, and
right-handed neutrinos, $\theta_{23}^R=\pi/4$  (in the notation of Sec.~\ref{sec:seesaw}) \cite{Plentinger:2007px}. Note also that in $M_D$ it is the 2nd column that is dominant. The PMNS mixing angles for
this texture set are given by\footnote{Note that future experimental bounds might afford it to consider corrections to the solar and atmospheric mixing angle stemming e.g. from the charged lepton sector.}
\begin{equation}
 \theta_{12}= \frac{\pi}{4}-\frac{\epsilon}{\sqrt{2}}+\mathcal{O}(\epsilon^2),\quad
\theta_{23}=\frac{\pi}{4}+\frac{\epsilon}{\sqrt{2}}+\mathcal{O}(\epsilon^2),\quad
\theta_{13}\simeq \frac{\epsilon^2}{2}.
\end{equation}
Note that this texture set has the interesting property that $\theta_{13}$ is suppressed by two powers of the Cabibbo angle. We have checked numerically  that this is a stable feature under variation of the lepton Yukawa couplings and mass terms. After rotating to the basis where $M_\ell$ is diagonal, the normalized neutrino Yukawa coupling matrix is to leading order in $\epsilon$ given by
\begin{equation}\label{eq:YDtexture}
Y_D=0.045\left(
\begin{matrix}
\epsilon^2& -1.7\epsilon^2 & - 0.28\epsilon^2\\
\epsilon & 0.71\epsilon & -0.71\epsilon\\
-0.28\epsilon^2 & -1 & -1
\end{matrix}
\right)\simeq
\frac{(m_\nu m^R_3)^\frac{1}{2}}{v\,\text{sin}\,\beta}
\left(
\begin{matrix}
 \epsilon^2 & \epsilon^2 & \epsilon^2\\
\epsilon & \epsilon & \epsilon\\
\epsilon^2 & 1 & 1
\end{matrix}
\right).
\end{equation}
where we have taken $m_\nu=5\times 10^{-2}\,\text{eV}$, $m^R_{3}=2.5\times 10^{12}\,\text{GeV}$ and
$\text{tan}\,\beta=10$. With this choice  the other heavy right-handed neutrino masses are $m_1^R=10^{11}\,\text{GeV}$ and $m_2^R=5\times 10^{11},\text{GeV}$. 
The dimensionless coefficient $0.045$ multiplying the matrix in (\ref{eq:YDtexture}) is set by the scales $m_\nu,m^R_3,$ and $v\,\text{sin}\beta$. Note that even after rotating to this basis, the factors multiplying within the matrix the powers of $\epsilon$ are all approximately of order one. 

\subsection{Complex Textures}\label{sec:CPviolating}
The reference list contains only real matrices and therefore describes only CP-conserving cases. We introduce
CP-violation for the different texture sets by multiplying each mass matrix element in the Lagrangian in
(\ref{eq:massterms}) by an individual random phase. This means that we modify for the texture set $\{M_\ell,M_D,M_R\}$ each mass matrix element to
\begin{equation}\label{eq:mapping}
(M_x)_{ij}\rightarrow (M_x')_{ij}=\text{exp}({\text{i}\alpha_{ij}^x})\cdot(M_x)_{ij}\quad(x=\ell,D,R),
\end{equation}
where we assume that the $9+9+6=24$ phases $\alpha^x_{ij}$ vary independently on a linear scale over the whole interval $\alpha^x_{ij}\in[0,2\pi[$. Notice that one can rotate away 10 of the 24 phases by phase-redefinitions, but we will not make use of this freedom here. Moreover, the mapping in (\ref{eq:mapping}) leaves the
moduli of the mass matrix elements invariant but changes the matrix $R$ in (\ref{eq:RMatrix}), the PMNS mixing
parameters, and the Dirac and Majorana phases $\delta,\phi_1,$ $\phi_2$. For each of the 72 real texture sets $\{M_\ell,M_D,M_R\}$ from the reference list we will thus obtain corresponding CP-violating texture sets
$\{M'_\ell,M_D',M_R'\}$. In the following, we will call a complex texture set obtained in this way from the texture No.~$i$ ($i=1,\dots, 72$) in the reference list a ``complex type-$i$ texture set''.

Among these complex texture sets, we further consider only those with $U_\text{PMNS}\simeq U_\text{HPS}$ (up to Majorana phases), i.e. we study only complex cases where the PMNS matrix remains of nearly tribimaximal
mixing form after inclusion of the CP-violating phases. Specifically, we require that the CP-violating textures
lead to solar and atmospheric mixing angles within the current $1\sigma$ errors in (\ref{eq:mixingangles}) while the reactor angle is very small and satisfies the constraint $\theta_{13}<5^\circ$ (cf.~Sec.~\ref{sec:CPconserving}).\footnote{Note that the PMNS angles $\theta_{12}$ and $\theta_{23}$ of the complex textures are therefore closer to the best-fit values (at $1\sigma$) than the real textures from the reference list (at $3\sigma$). In addition, we concentrate on small values of $\theta_{13}$ since this limit is interesting for flavor models and is near the best-fit value.} Moreover, we demand that the complex textures approximately
reproduce (in our examples up to relative factors of $1.5$) the mass eigenvalues in (\ref{eq:chargedleptonmasses}), (\ref{eq:neutrinomasses}), and (\ref{eq:leptonmasses}), of the original texture set from the reference list they have been derived from. In this way, each of the 72 real examples from the reference list serves as a ``parent'' to a whole class of complex ``daughter'' textures that have mass ratios that are similar to those of the real
parent texture set and also show all nearly tribimaximal lepton mixing, but they vary strongly in the CP-violating phases appearing in the Lagrangian.

In addition, to ensure perturbativity of the Higgs sector, we will require that the Dirac neutrino matrices always satisfy 
\begin{equation}\label{eq:perturbativity}
|Y_{3}^D|^2/(4\pi)<0.3,
\end{equation}
where $|Y_{3}^D|^2$ is the absolute value of the largest eigenvalue of $Y_D^\dagger Y_D$, which is usually set by the
largest Yukawa coupling in $Y_D$. Note that this perturbativity constraint becomes particularly important for large $m^R_3 \gtrsim 10^{14}\,\text{GeV}$.

In the parametrization (\ref{eq:CI}), the complex textures will in general lead to a complex $R$ matrix
and, therefore, to nonzero parameters $y_i$. Consider, as an example, this parametrization for a random complex texture set that is obtained from texture set No.~1 by introducing CP-phases via the mapping in (\ref{eq:mapping}).
 Similar to \cite{Deppisch:2005rv}, we approximate
$M^\text{diag}_R\approx m^R_3 \,\text{diag}(0,0,1)$ and
$M_\text{eff}^\text{diag}\approx
m_\nu\,\text{diag}(0,\epsilon,1)$. Going to exact tribimaximal mixing
$U_\text{PMNS}\simeq U_\text{HPS}$, we thus obtain by comparison with (\ref{eq:YD}) that
\begin{equation}\label{eq:approxYD}
 (Y_D)_{ij}\simeq\frac{(m_\nu m^R_3)^\frac{1}{2}}{v\,\text{sin}\,\beta}
c_2\delta_{i3}\left[s_1e^{\text{i}\phi_2}\sqrt{\epsilon}
(U_\text{HPS})_{j2}+c_1(U_\text{HPS})_{j3}\right].
\end{equation}
From this we conclude that $c_1c_2\simeq\sqrt{2}$ and $2s_1c_2\sqrt{\epsilon}/\sqrt{3}\approx 0$. Therefore, $s_1\approx 0$ and comparison with (\ref{eq:YDtexture}) leads to small values $x_1,y_1\approx 0$. This implies that
$c_1\simeq 1$ and $c_2\simeq\sqrt{2}$. In contrast to this, there are no obvious strong constraints from (\ref{eq:approxYD}) on the possible values of $x_2,x_3,$ and $y_3$ (apart from perturbativity arguments). We therefore
roughly take an average value $\text{sin}\,x_2\simeq 2/\pi$. For the complex textures, we then expect from $c_2\simeq\sqrt{2}$ also that $\text{sin}\,x_2\,\text{sinh}\,y_2\simeq \sqrt{2}$. As a result, we
arrive crudely at a typical value
\begin{equation}\label{eq:y2}
 y_2\simeq\text{arcsinh}\big(\pi/\sqrt{2}\big)\simeq\mathcal{O}(1),
\end{equation}
where the factor $1/\sqrt{2}$ in the argument is a consequence of the approximation by tribimaximal mixing. From (\ref{eq:y2}), we expect in the distribution of the $y_i$ a clustering of $|y_2|$ at values $\simeq\mathcal{O}(1)$. Note that perturbativity constrains the $y_i$ not to become much larger than one.

\begin{figure}
\begin{center}
\begin{tabular}{ccc}
\includegraphics*[height=3.5cm]{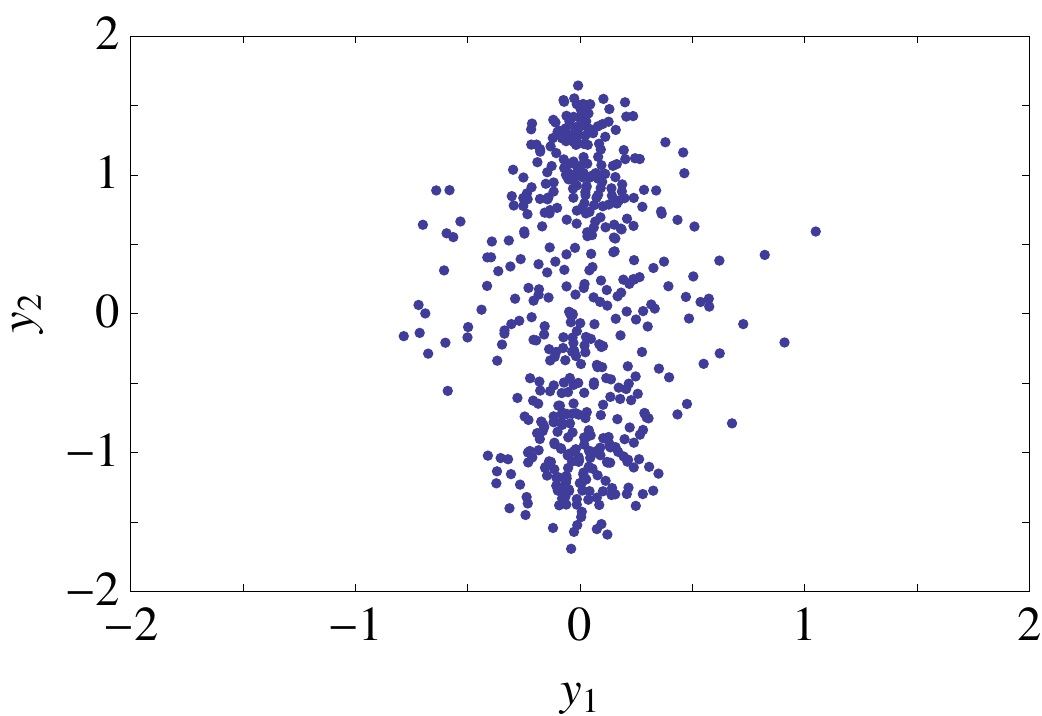}&
\includegraphics*[height=3.5cm]{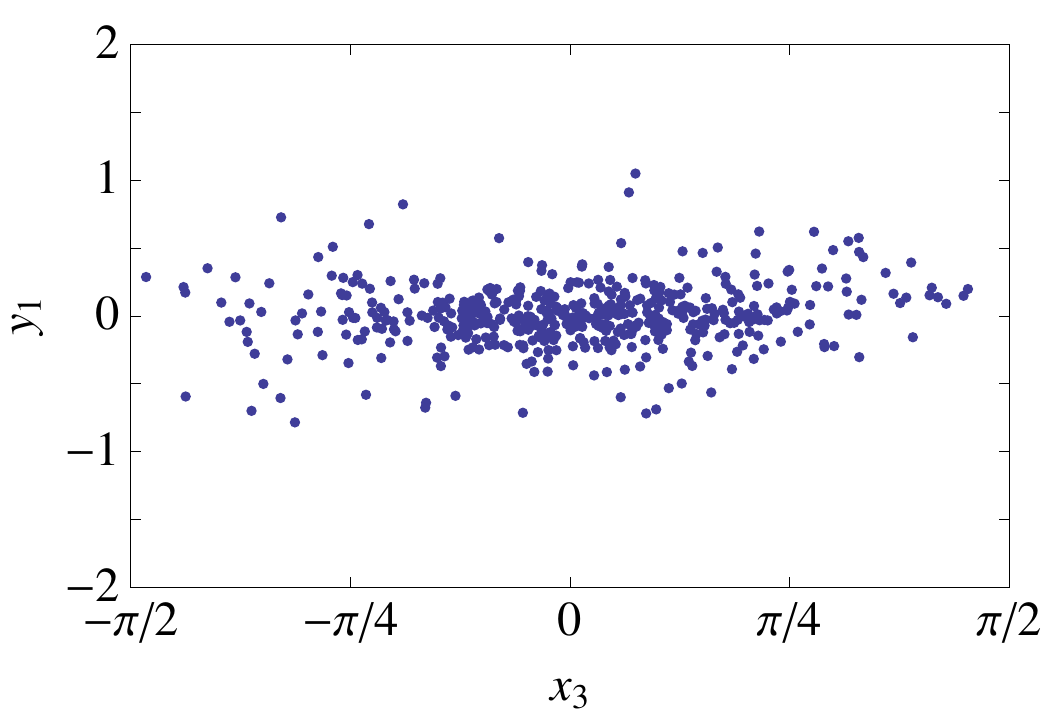}&
\includegraphics*[height=3.5cm]{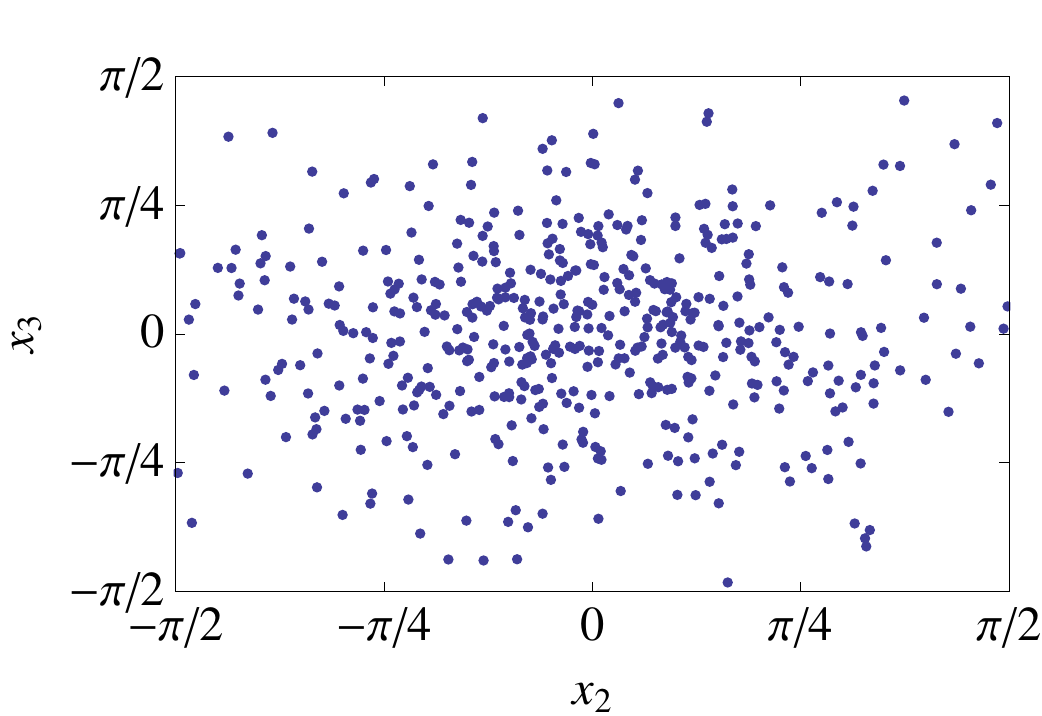}
\end{tabular}
\caption{Distribution of the parameters $x_i$
  and $y_i$ [cf.~(\ref{eq:RMatrix})] for 500 complex type-1
  textures with a heaviest right-handed neutrino mass $m_3^R=2.5\times 10^{12}\,\text{GeV}$.}\label{fig:Rparameters}
\end{center}
\end{figure}

In Fig.~\ref{fig:Rparameters}, we show the distribution of the parameters $x_i$ and $y_i$ belonging to 500 complex type-1 textures, which all satisfy the perturbativity constraints in (\ref{eq:perturbativity}) for $m_3^R=2.5\times 10^{12}\,\text{GeV}$.
 The complex type-1 textures are all obtained from the real  texture set No.~1 in the reference list by scattering only the phases in the Lagrangian. Fig.~\ref{fig:Rparameters} displays, as explained above, only the parameters $x_i$ and $y_i$ of the complex textures which are consistent with nearly tribimaximal lepton mixing at the $1\sigma$ level and reproduce, up to relative factors of 1.5, the lepton mass ratios given in (\ref{eq:chargedleptonmasses}),
(\ref{eq:neutrinomasses}) and (\ref{eq:leptonmasses}).

We observe that $|y_2|$ has, as estimated above, indeed a weak clustering at values $\approx 1$, while $y_1$ stays mostly at values $|y_1|\lesssim 0.3$, whereas $x_2$ and $x_3$ essentially vary over the whole interval $[-\pi/2,\pi/2]$. In addition, $|y_3|$ can become $\simeq\mathcal{O}(1)$ without any clear preference for certain values in this range. As expected, $x_1$ is small, with values $|x_1|\lesssim \pi/4$. We will later see that the
distribution of parameters allows to generate easily large Dirac and Majorana CP-violation phases in the low-energy theory.

The parametrization of $Y_D$ in terms of the $R$ matrix (\ref{eq:RMatrix}) makes it obvious that the inclusion of general CP-violating phases via (\ref{eq:mapping}) leads to an increase of the LFV rates. On the other hand, since the moduli of the Yukawa couplings are held fixed in this mapping one may expect that the increase in the LFV branching ratios still remains moderate. From (\ref{eq:y2}), we can estimate that the complex textures
will usually have LFV-rates that are roughly by a factor $(\pi/\sqrt{2})^4\simeq \mathcal{O}(10)$ larger
than for the real textures.

\section{Results for LFV Rates}\label{sec:results}

In this section, we present our results for the LFV-rates of the textures with nearly tribimaximal lepton mixing. First, we consider the LFV-rates for our reference list of real textures, i.e.~the CP-conserving case. Then, we
turn to the complex textures generalizing the reference list by including in the most general way random CP-violating phases at the Lagrangian level. As input parameters for the LFV-rates we take the GUT scale $M_X=2.5\times 10^{16}\,\text{GeV}$ and an effective neutrino mass scale $m_\nu=5\times 10^{-2}\,\text{GeV}$. Throughout, the light
neutrinos have the normal hierarchical spectrum given in (\ref{eq:neutrinomasses}). Unless stated otherwise,
we will always refer to the mSUGRA benchmark scenario SPS1a' \cite{AguilarSaavedra:2005pw}. The scenario SPS1a' has a
universal gaugino mass $m_{1/2}=250\,\text{GeV}$ and a universal scalar mass $m_0=70\,\text{GeV}$ at the GUT scale, $\text{tan}\beta=10$, a positive sign $\text{sign}(\mu)=+$ of the Higgs mixing parameter $\mu$, and a universal trilinear coupling parameter $A_0=-300$~GeV.

\subsection{Real Textures}
\begin{figure}[t]
\begin{center}
\begin{tabular}{cc}
\includegraphics[width=0.49\textwidth]{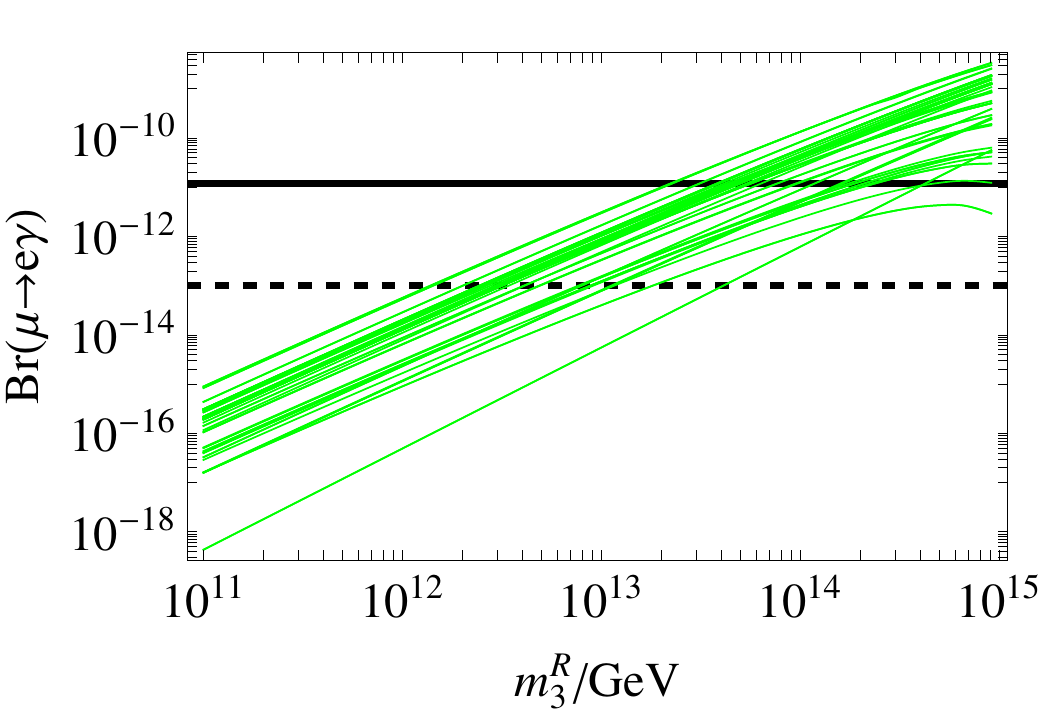}&
\includegraphics[width=0.49\textwidth]{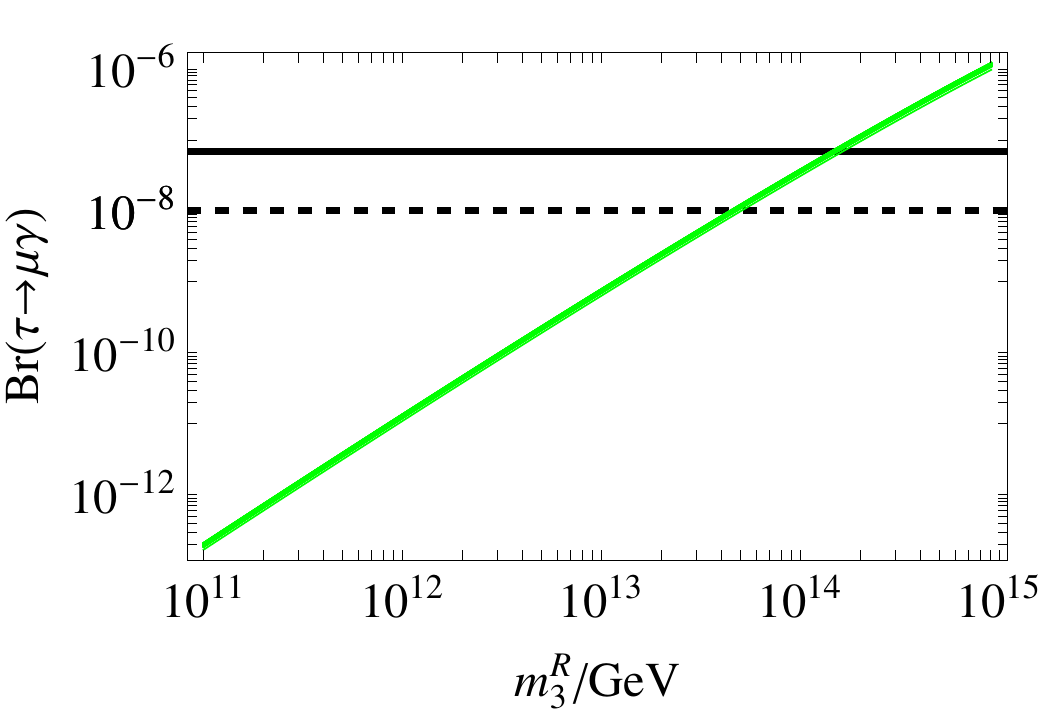}\\
\includegraphics[width=0.49\textwidth]{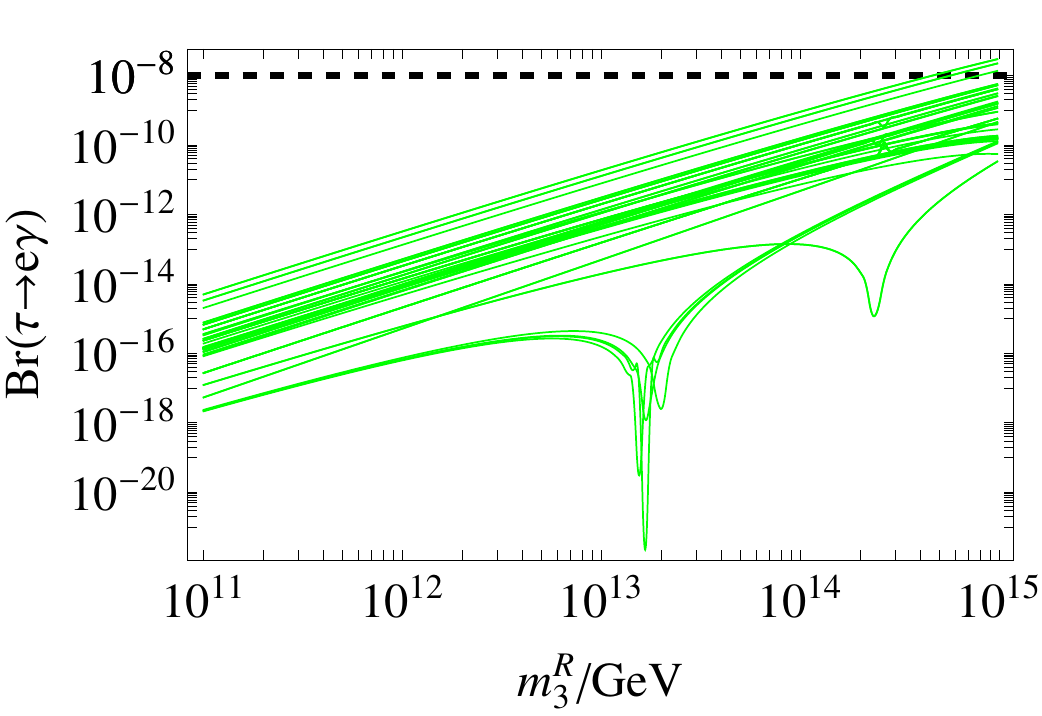}&
\includegraphics[width=0.49\textwidth]{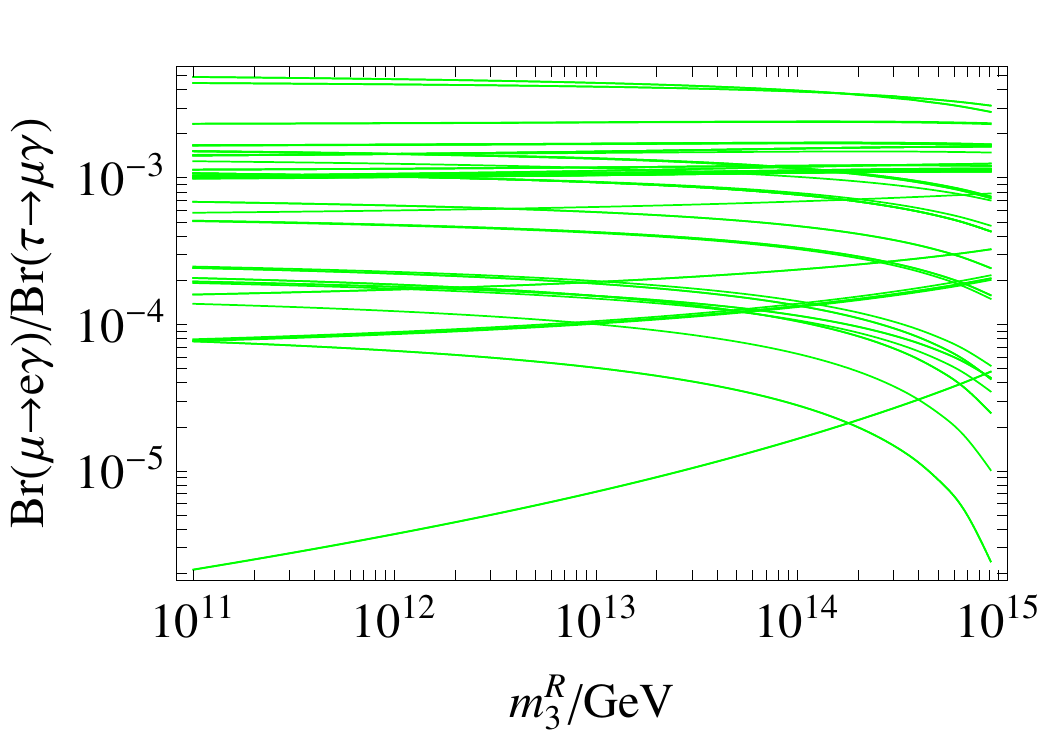}
\end{tabular}
\caption{$\text{Br}(\mu\rightarrow e\gamma)$ (top left), $\text{Br}(\tau\to\mu\gamma)$ (top right), $\text{Br}(\tau\to e\gamma)$ (bottom left), and ${\text{Br}(\mu\rightarrow e\gamma)}/{\text{Br}(\tau\to\mu\gamma)}$ (bottom right) as a function of the heaviest right-handed neutrino mass $m_3^R$ for the reference list of 72 real texture sets in the SUSY scenario SPS1a'. The solid (dashed) lines represent the current (future) experimental bounds on the respective LFV branching ratio.}\label{fig:72brs}
\end{center}
\end{figure}
\begin{figure}[t!]
\centering
\includegraphics[width=0.6\textwidth]{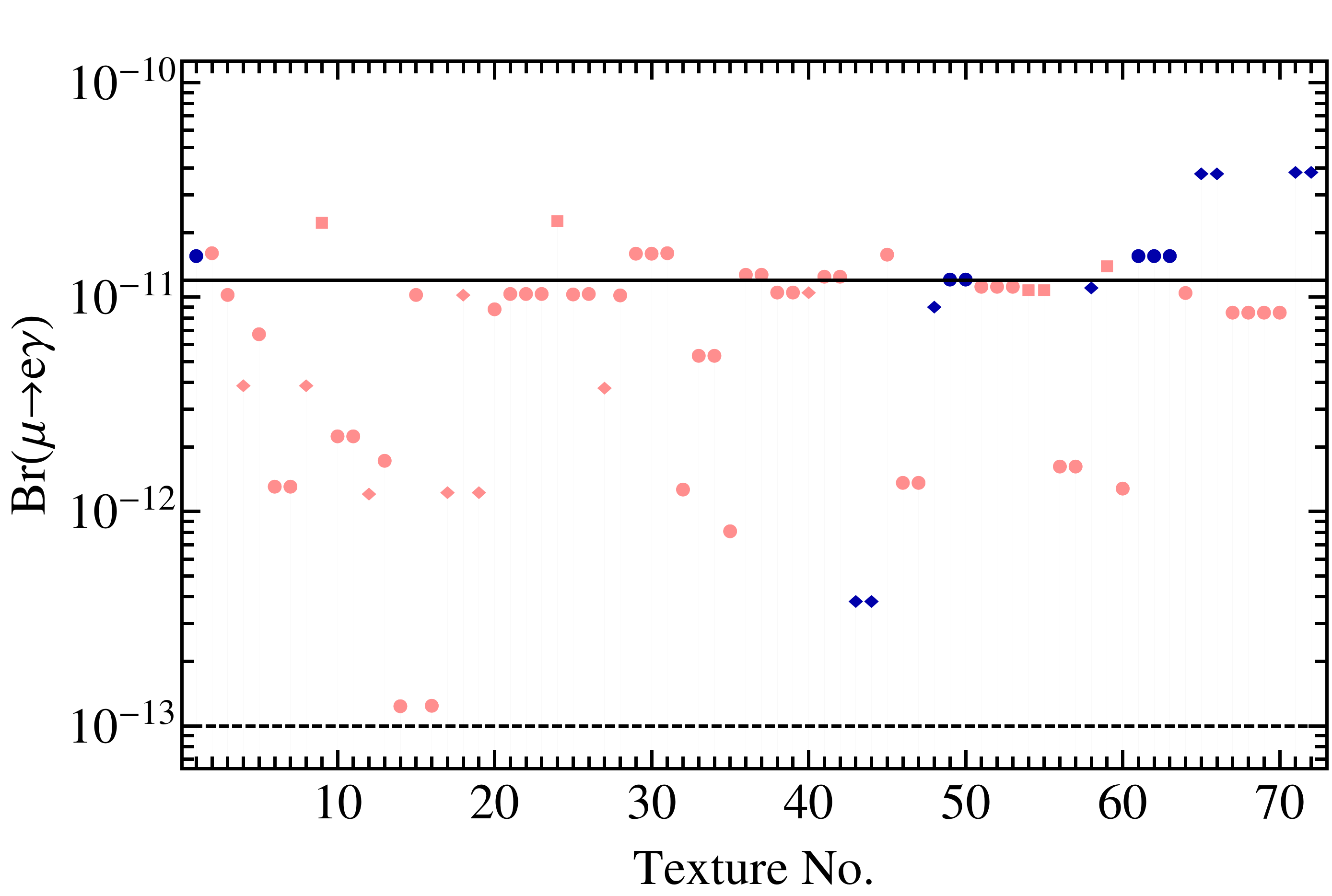}
\caption{$\text{Br}(\mu\to e\gamma)$ for all 72 real texture sets from the
  reference list for a fixed heaviest right-handed neutrino mass
  $m_3^R=4.5\times 10^{13}$~GeV (corresponding to
  $\text{Br}(\tau\to\mu\gamma)\approx 10^{-8}$ for all texture sets) at the
  SUSY benchmark point SPS1a'. The hierarchy in the heavy right-handed
  neutrino spectrum of the respective texture set is denoted as dark blue (strictly hierarchical) and light red (degeneracy between the two heaviest right-handed neutrinos). The symbol shape denotes the hierarchy of the Dirac neutrino mass eigenvalues: diamond (normal hierarchy), circle ($m_2^D$ is largest eigenvalue) and square ($m_1^D$ is largest eigenvalue). The solid (dashed) line represents the current (future) experimental bound.}\label{fig:mue_scenario}
\end{figure}

Let us consider first the LFV decay rates for the 72 texture sets in our reference list. All the textures in the reference list describe real lepton mass terms and are, thus, CP-conserving. Fig.~\ref{fig:72brs} shows the LFV branching ratios $\text{Br}(\mu\rightarrow e\gamma)$, $\text{Br}(\tau\to\mu\gamma)$, $\text{Br}(\tau\to e\gamma)$, as well as the ratio $\text{Br}(\mu\to e\gamma)/\text{Br}(\tau\to e\gamma)$ for the 72 texture sets as a function of the heaviest right-handed Majorana neutrino mass $m_3^R$. While $\text{Br}(\mu\rightarrow e\gamma)$ varies for fixed $m_3^R$ by at least one order of magnitude for different texture sets, $\text{Br}(\tau\to\mu\gamma)$ changes, in comparison, hardly at all. As these two LFV rates scale similarly for small values of $m_3^R\ll M_X$, $\text{Br}(l_i\rightarrow l_j\gamma)\propto (m_3^R)^2$, in most of the textures, the ratio $\text{Br}(\mu\to e\gamma)/\text{Br}(\tau\to \mu\gamma)$ is approximately independent of $m_3^R$ in the shown regime. Only for a handful of textures, the simple scaling is violated in $\text{Br}(\mu\rightarrow e\gamma)$, and $\text{Br}(\mu\to e\gamma)/\text{Br}(\tau\to \mu\gamma)$ anomalously increases with increasing $m_3^R$. This occurs most pronouncedly in textures No.~14 and 16 which exhibit a strongly suppressed $\mu\to e\gamma$ rate (cf. Fig.~\ref{fig:mue_scenario}). The $\tau\rightarrow e\gamma$ decay rate varies at least by two orders of magnitude, and for some textures exhibits a pronounced kink where the rate is drastically suppressed due to accidental cancellations. All texture sets from the reference list can have $m_3^R$ as large as  $\sim 2\times 10^{13}\,\text{GeV}$ in agreement with the current bound on $\text{Br}(\mu\rightarrow e\gamma)\lesssim 10^{-11}$. Future non-observation of $\mu\rightarrow e\gamma$ at the upcoming PSI experiment, however, would imply for almost all texture sets that $m_3^R\lesssim 10^{13}\,\text{GeV}$. While $\text{Br}(\tau\to\mu\gamma)$ could be measured in the future for $m_3^R\gtrsim 5\times 10^{13}\,\text{GeV}$, a measurement of $\tau\rightarrow e\gamma$ is out of reach for $m_3^R\lesssim 10^{15}\,\text{GeV}$ for most of the textures.

Next, let us try to gain some rough understanding of the relative size of the LFV rates in terms of the lepton mass hierarchies and the radiative effects of the right-handed neutrinos on the slepton mass matrix. In Fig.~\ref{fig:mue_scenario} we show $\text{Br}(\mu\to e\gamma)$ for the 72 real texture sets from the reference list for a fixed right-handed neutrino mass $m_3^R=4.5\times 10^{13}$~GeV. It should be noted that this choice for $m_3^R$ is different from the value used in the rest of the paper and was chosen here to have $\text{Br}(\tau\to\mu\gamma)$ near the future experimental sensitivity. Fig.~\ref{fig:mue_scenario} demonstrates that if $\text{Br}(\tau\to\mu\gamma)>10^{-8}$ is observed in the future, all 72 texture sets can be probed at the SUSY benchmark point SPS1a' by measuring $\mu\to e\gamma$. Assuming a sufficiently exact and independent measurement of the SUSY mass spectrum, e.g. at the LHC or by correlating the LFV decay rates with the anomalous dipole moment of the muon~\cite{Deppisch:2002vz}, the observation of $\tau\to\mu\gamma$ would provide an unambiguous determination of the heaviest right-handed neutrino mass $m_3^R>4.5\times 10^{13}$~GeV, and an observation of $\mu\to e\gamma$ would then constrain the number of viable texture sets, cf.~Fig.~\ref{fig:mue_scenario}. Whereas a unique identification of a particular texture set seems in general difficult, due to the fact that several texture sets have coincident or similar $\mu\to e\gamma$ rates, such an optimal observational picture would allow a much deeper
insight into the possible leptonic flavor structure. An observable rate of $\tau\to e\gamma$ at future experiments would require a very high right-handed neutrino mass scale $m_3^R > 4\times 10^{14}$~GeV, which is incompatible with the non-observation of $\tau\to\mu\gamma$. 

\subsection{Complex Textures}
\begin{figure}[t]
\begin{center}
\includegraphics[width=0.6\textwidth]{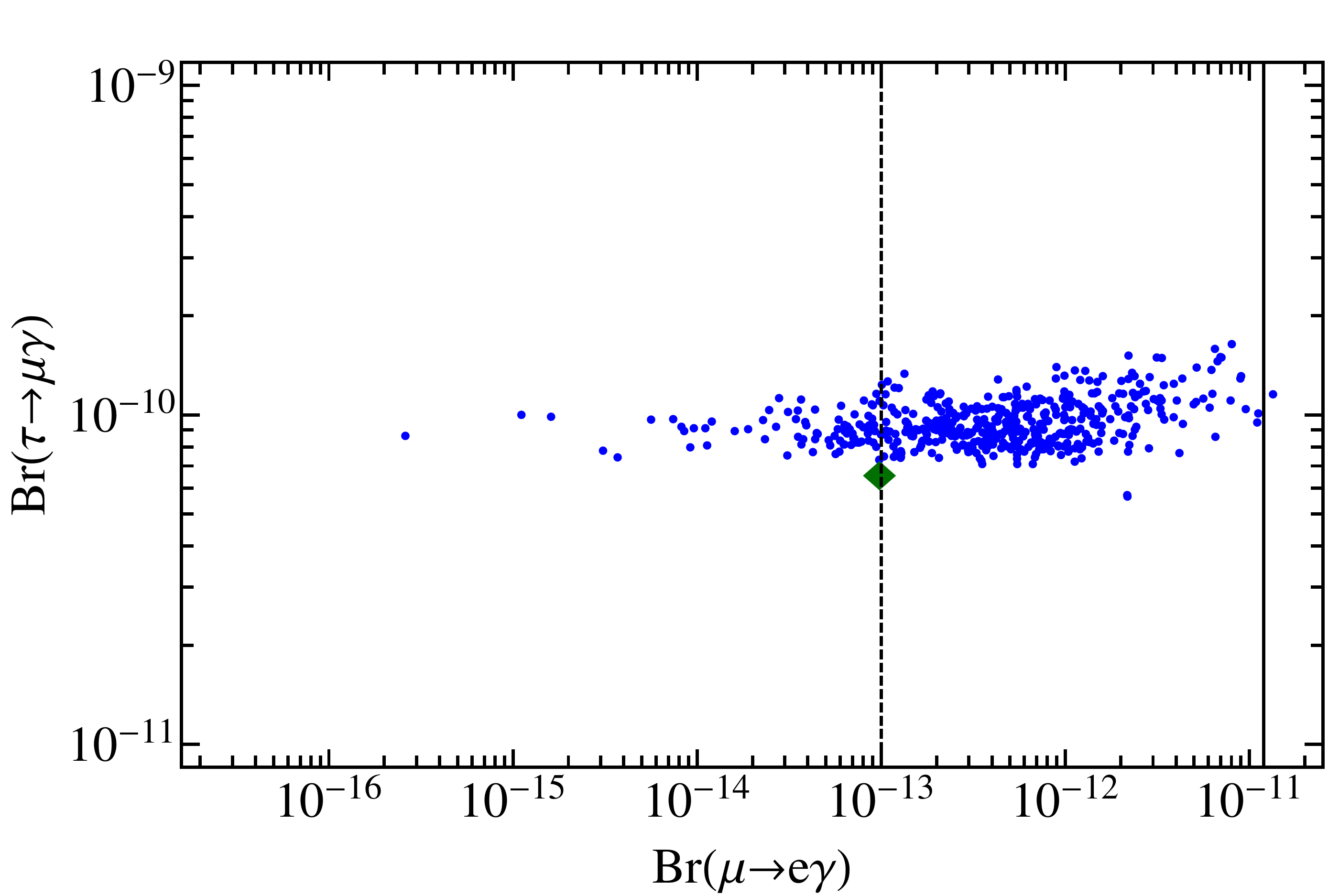}
\caption{$\text{Br}(\mu\to e\gamma)$ and $\text{Br}(\tau\to\mu\gamma)$
  for the texture set No.~1 with $m_3^R=2.5\times 10^{12}$~GeV in the
  SUSY scenario SPS1a'. The green diamond shows the branching ratios
  for the CP-conserving case while the blue points denote the
  branching ratios for 500 random complex type-1 texture sets. The solid (dashed) line represents the current (future) experimental bound on $\text{Br}(\mu\to e\gamma)$.}\label{fig:MRScan_Texture_1}
\end{center}
\end{figure}
Now, let us turn to the complex case, which we will study for a specific example texture set. In what follows, we will focus on texture set No.~1 from the reference list discussed in Sec.~\ref{sec:textures}, but our results are at least qualitatively representative for the complete reference list. Figure \ref{fig:MRScan_Texture_1} shows $\text{Br}(\mu\to e\gamma)$ vs. $\text{Br}(\tau\to \mu\gamma)$ for 500 complex type-1 textures in the
SUSY scenario SPS1a'. The 500 complex examples are generated from the real texture (denoted by the green diamond in
Figure~\ref{fig:MRScan_Texture_1}) in the reference list as explained in Sec.~\ref{sec:CPviolating}. They give all nearly tribimaximal lepton mixing at the $1\sigma$ level and exhibit, up to relative factors of 1.5, the lepton mass ratios listed in (\ref{eq:chargedleptonmasses}), (\ref{eq:neutrinomasses}), and (\ref{eq:leptonmasses}). We observe from Fig.~\ref{fig:MRScan_Texture_1} that the complex textures have maximal values for $\text{Br}(\mu\rightarrow e\gamma)$ up to a factor $\sim 200$ larger than for the corresponding real case. In exceptional cases, $\text{Br}(\mu\rightarrow e\gamma)$ can be smaller by three orders of magnitude than in the CP-conserving case. On the other hand, $\text{Br}(\tau\to \mu\gamma)$ only varies by a factor of $\sim 3$ due to arbitrary CP phases. Because of the potential increase or decrease of $\text{Br}(\tau\to \mu\gamma)$ by 2-3 orders of magnitude
for nonzero CP phases, it seems generally not possible to discriminate the complex texture sets through the observation of LFV rates. For completeness and consistency, we have checked that the electron electric dipole moments for the complex textures are several orders of magnitude below current bounds, as expected in general mSUGRA seesaw models with a mild right-handed neutrino mass hierarchy. \cite{Ellis:2002xg}.

\begin{figure}[t]
\begin{center}
\begin{tabular}{cc}
\includegraphics*[height=5cm]{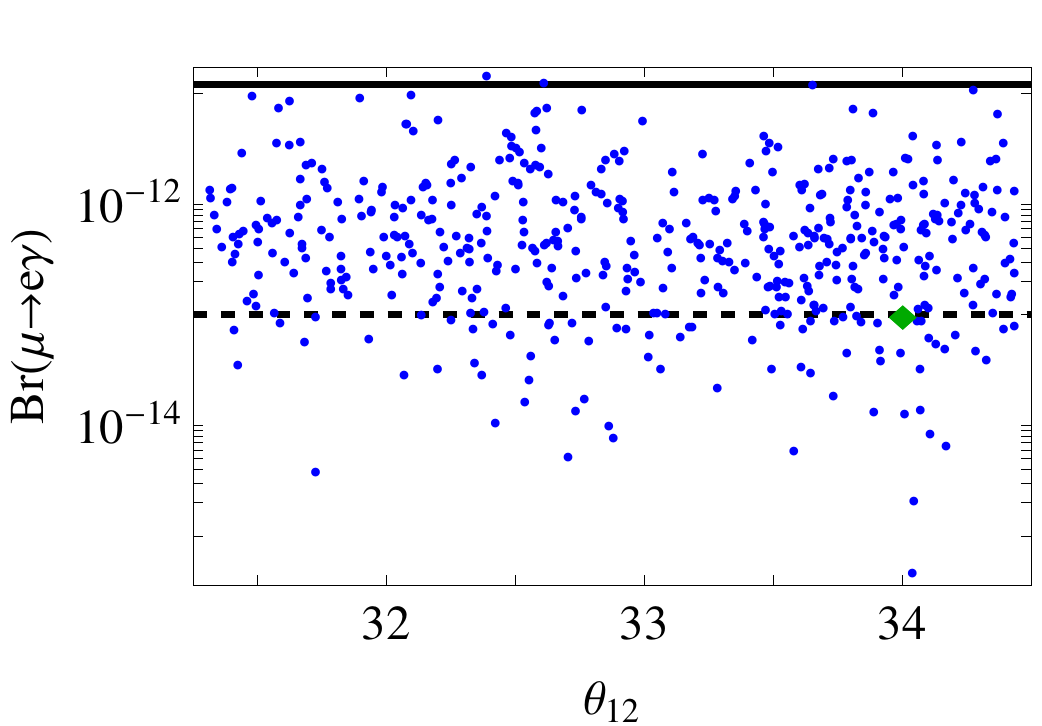}&
\includegraphics*[height=5cm]{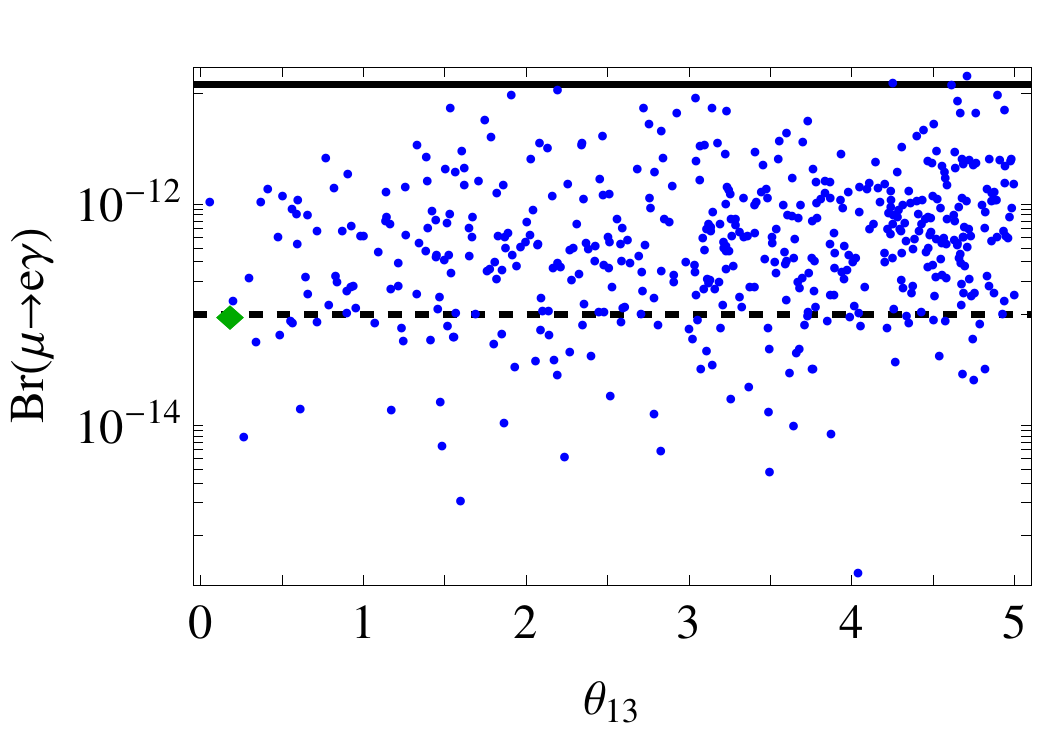}\\
\includegraphics*[height=5cm]{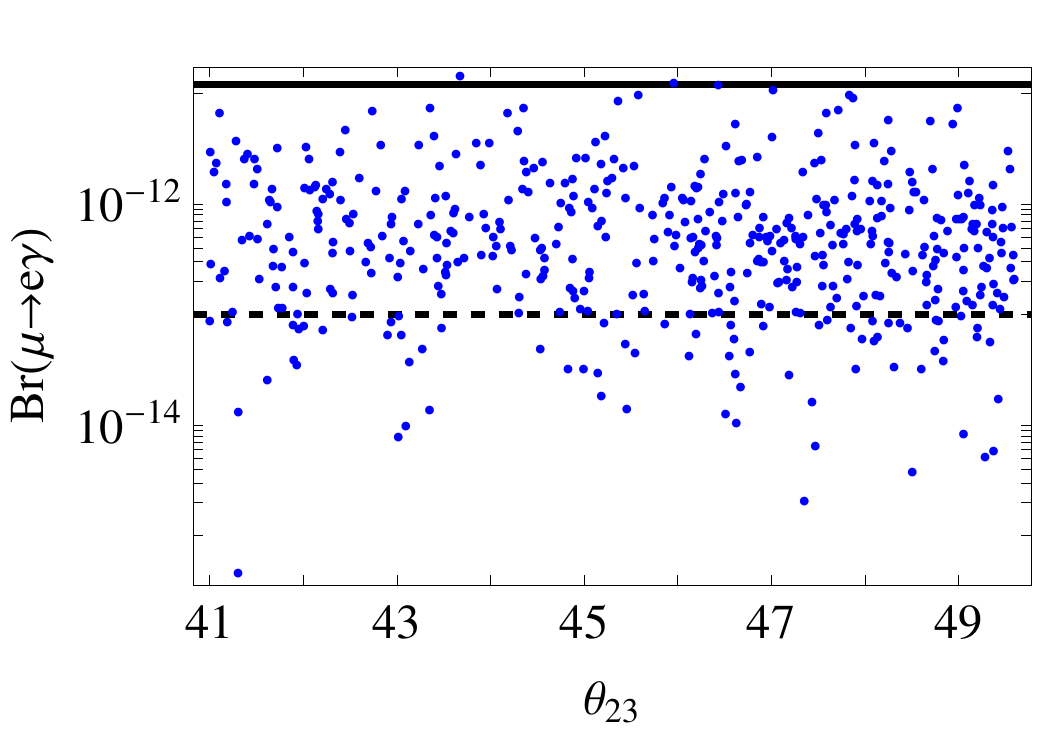}&
\includegraphics*[height=5cm]{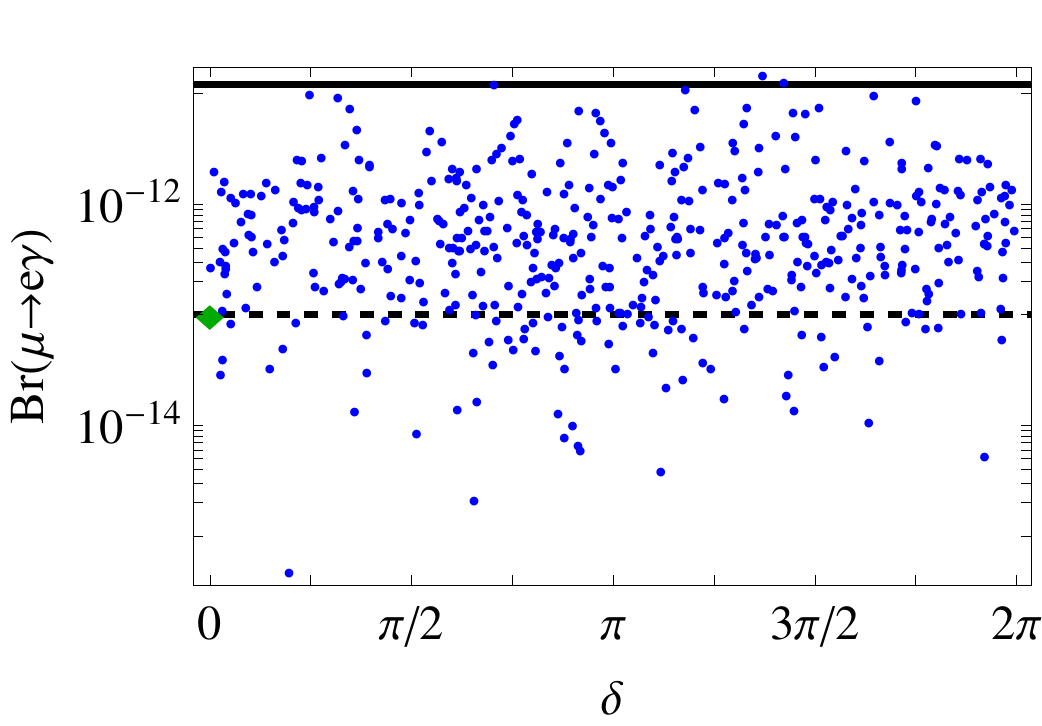}
\end{tabular}
\caption{$\text{Br}(\mu\rightarrow e\gamma)$ vs.~the PMNS mixing angles and Dirac CP-phase for the complex textures scanned in Fig.~\ref{fig:Rparameters}. The green diamond shows the values for the CP conserving case. The solid (dashed) line represents the current (future) experimental bound.}\label{fig:PMNSparameters}
\end{center}
\end{figure}

Next, let us consider the LFV branching ratios as a function of the low-energy PMNS mixing parameters. Fig.~\ref{fig:PMNSparameters} summarizes the nearly tribimaximal PMNS mixing parameters and Dirac phase belonging to the complex textures scanned in Figs.~\ref{fig:Rparameters}. We see that the solar and atmospheric angle $\theta_{12}$ and $\theta_{23}$ populate the complete allowed ranges $\theta_{12}=33^\circ\pm 1.5^\circ$ and $\theta_{23}=45^\circ\pm 4^\circ$ with no clear preference for any value. The same applies to the Majorana
and Dirac CP-violation phases $\phi_1,\phi_2$ and $\delta$, which can take any value in the intervals $\phi_{1,2}\in [0,\pi[$ and $\delta\in [0,2\pi[$ (shown is only the Dirac CP-violation phase $\delta$). In particular, the complex textures do not prefer large or small values for the phases, even though they have been generated by starting out
with a real texture set. Concerning the reactor angle $\theta_{13}$, however, we find that the presence of nonzero phases drives $\theta_{13}$ away from very small values $\theta_{13}\lesssim 1^\circ$, to larger values up to $5^\circ$, where we have set our selection cutoff, are preferred. This may be related to the way in which a small reactor angle can emerge for the textures No.~1 in the reference list and would therefore be the result of a possible underlying high-energy theory of flavor.

It has been pointed out previously that there may be a correlation between $\text{Br}(\mu\rightarrow e\gamma)$ and the value of the reactor angle $\theta_{13}$ for a special choice of parameters \cite{Masiero:2004js,Antusch:2006vw}. Fig.~\ref{fig:PMNSparameters}, however, does not exhibit any clear correlation between $\text{Br}(\mu\rightarrow e\gamma)$ and any of the PMNS mixing angles and phases. In particular, $\text{Br}(\mu\to e\gamma)$ varies by at least four orders of magnitude over the interval $\theta_{13}\in[0,5^\circ]$.

\begin{figure}[t]
\centering
\includegraphics[width=0.49\textwidth]{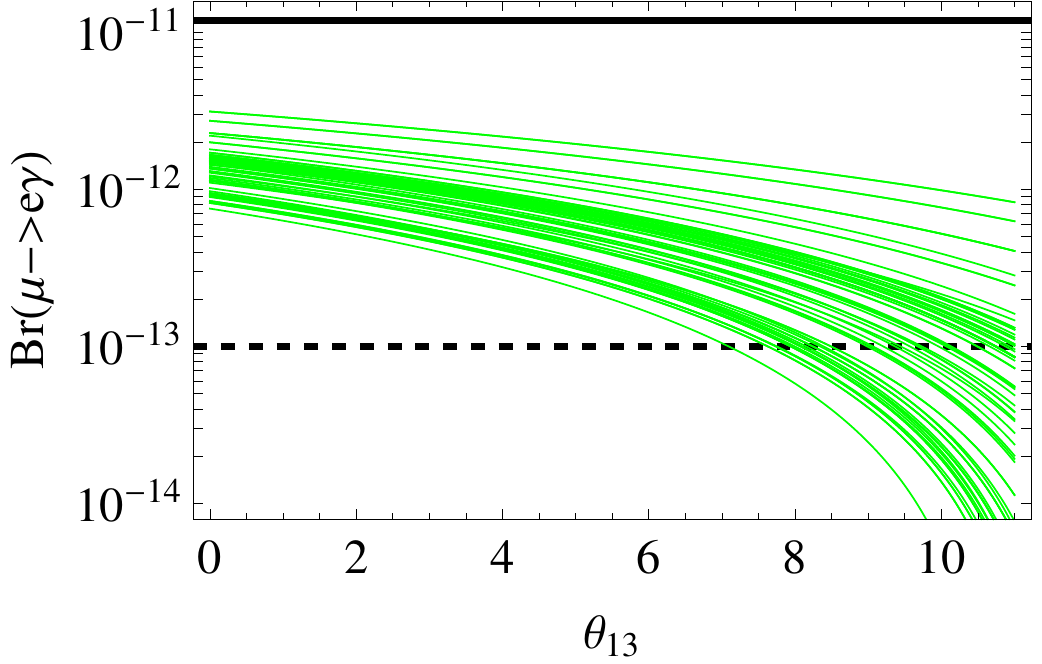}
\includegraphics[width=0.49\textwidth]{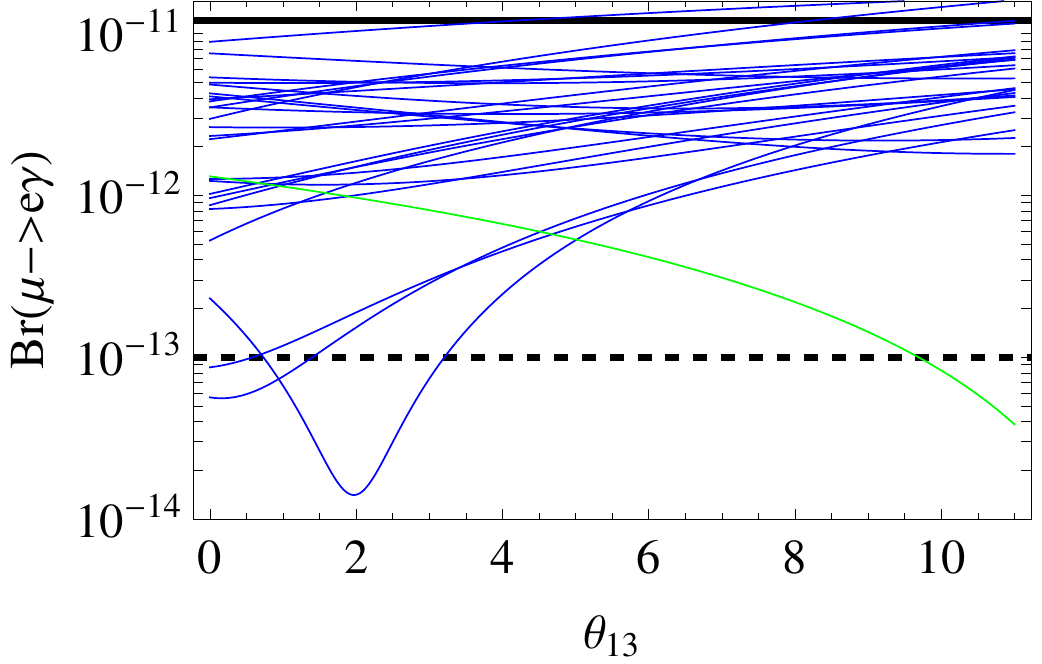}
\caption{$\text{Br}(\mu\to e\gamma)$ as a function of $\theta_{13}$ for the reference list of 72 real texture sets (left) and
  for 25 random complex type-1 texture sets (right) in the SUSY scenario SPS1a'. The 25 blue (dark) curves for texture No.~1 corresponding to a random choice of phases in the Yukawa couplings, while the green (light) curve corresponds to the CP-conserving case. The solid (dashed) line represents the current (future) experimental bound.}\label{fig:72brst13}
\end{figure}
Figure \ref{fig:72brst13} demonstrates the lack of a universal correlation between $\text{Br}(\mu\to e\gamma)$ and the value of the reactor angle $\theta_{13}$ in more detail. The left plot shows $\text{Br}(\mu\to e\gamma)$ as a function of $\theta_{13}$ for all 72 real texture sets, whereas the right plot shows the same dependence for 25 random complex type-1 texture sets. The plots were generated by calculating the $R$ matrix, cf. (\ref{eq:RMatrix}), and heavy right-handed neutrino masses in the respective texture (real or complex) and introducing a variable reactor angle through the $U_\text{PMNS}$ matrix in (\ref{eq:CI}), while keeping all other terms fixed. The neutrino Yukawa coupling matrix $Y_D$ is then used to calculate $\text{Br}(\mu\to e\gamma)$. While the real texture sets exhibit a qualitatively similar dependence on $\theta_{13}$, Fig.~\ref{fig:72brst13} (right) clearly proves that there is no general correlation between mSUGRA LFV rates and the neutrino mixing angle $\theta_{13}$ due to the unknown high-energy parameters as long as only the low-energy neutrino parameters are known. We have also checked other complex textures from the reference list with qualitatively similar results as for the complex type-1 texture set. An analysis supporting this result but is based on a parameter scan of the R matrix, can be found in \cite{Casas:2010wm}. However, particular models can exhibit indeed a $\theta_{13}$ dependence or correlation among low-energy phases \cite{Plentinger2010}.

\begin{figure}[t]
\begin{center}
\includegraphics[width=0.45\textwidth]{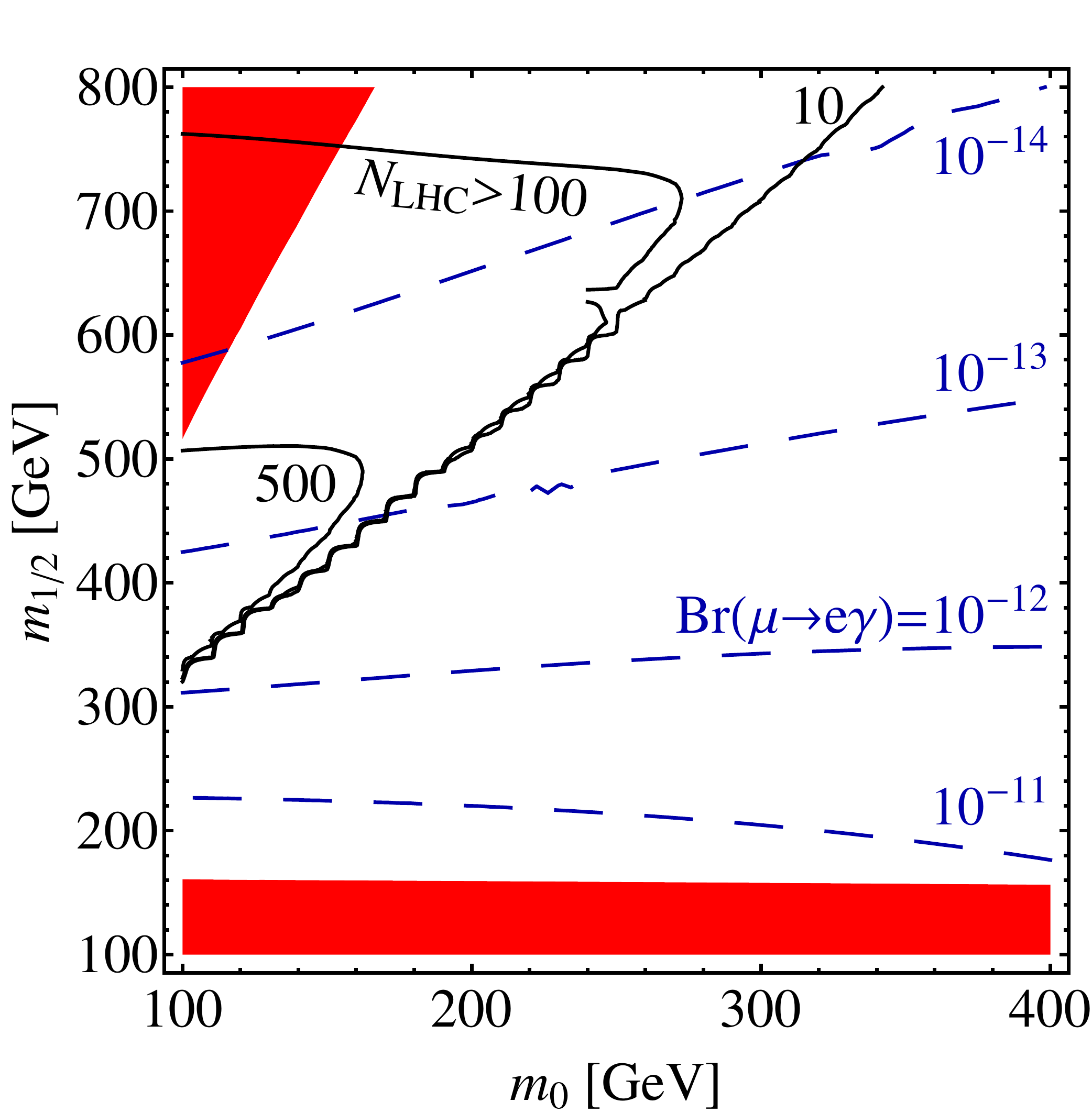}
\includegraphics[width=0.45\textwidth]{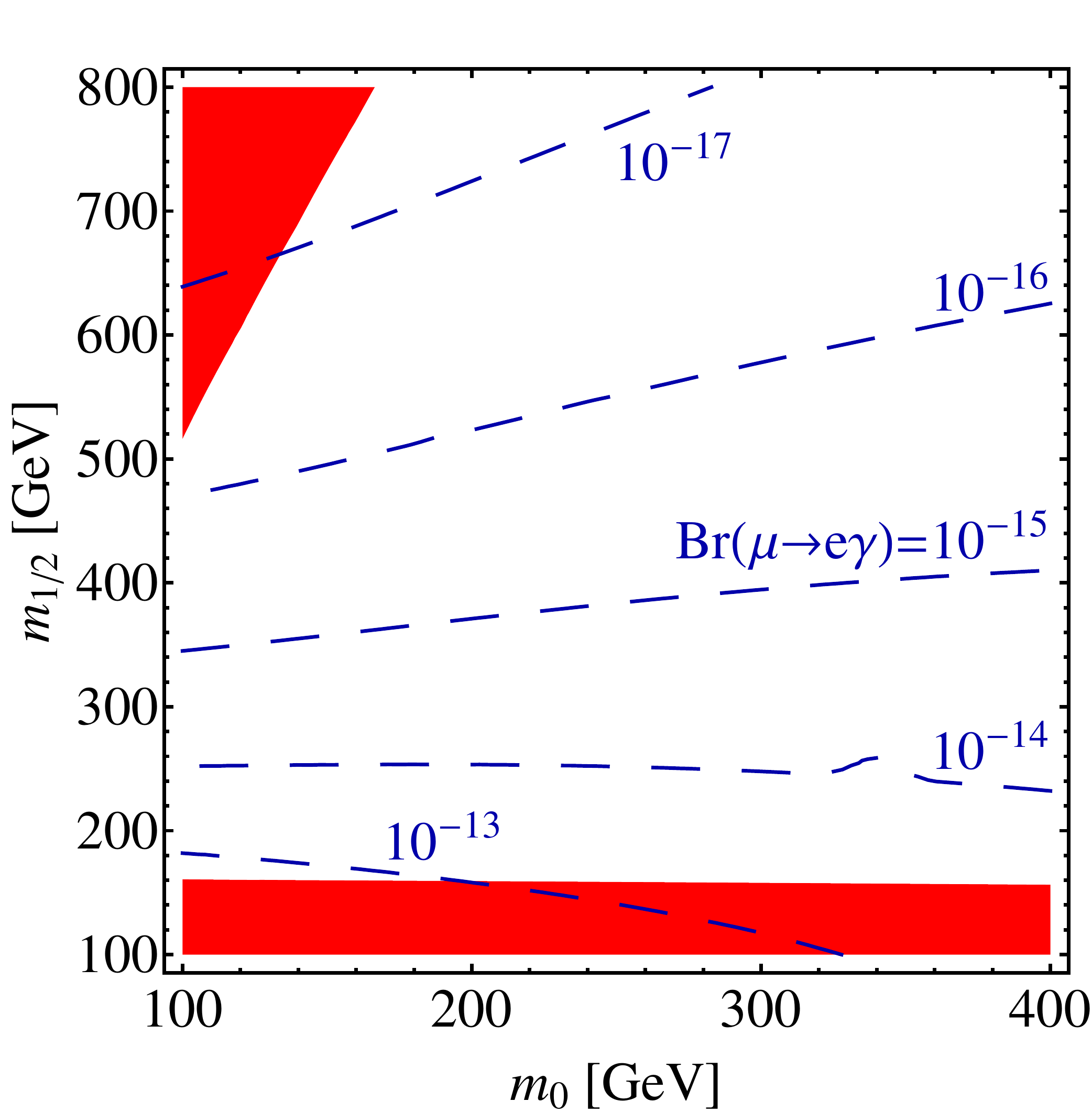}
\caption{Contours of $\text{Br}(\mu\to e\gamma)$ and the number of LFV
  events $N(pp \to\tilde\chi_2^0 + X \to e\mu\tilde\chi_1^0+X)$ at the
  LHC ($L=100 \text{ fb}^{-1}$) in the $(m_0-m_{1/2})$ parameter plane. The
  other mSUGRA parameters are chosen such that $A_0=-300$~GeV,
  $\tan\beta=10$ and $\mu>0$. On the left (right) side the neutrino sector is given as in
  texture 1, with the CP phases chosen such that maximal (minimal) 
 $\text{Br}(\mu\to e\gamma)$
is achieved for fixed $m^R_3 = 2.5 \times 10^{12}$~GeV.
 The dark (red) areas are excluded by direct SUSY searches.}\label{fig:lfv_max}
\end{center}
\end{figure}

Lepton flavor violating rates for decays of supersymmetric particles in the context of seesaw models with real parameters have already been discussed in the literature \cite{Raidal:2008jk,delAguila:2008iz,Hirsch:2008dy,Esteves:2009vg}. Therefore, we concentrate here on the effect of phases. In Fig.~\ref{fig:lfv_max}, we show for two specific complex examples belonging to the texture set No.~1 in the reference list the contours of $\text{Br}(\mu\to e\gamma)$ and the number of LFV events $N(pp \to\tilde\chi_2^0 + X \to e\mu\tilde\chi_1^0 + X)$ at the LHC ($L=100 \text{ fb}^{-1}$) in the $(m_0-m_{1/2})$ parameter plane. We consider those textures that lead (among the complex textures in Fig.~\ref{fig:MRScan_Texture_1}) to the highest  and lowest  rate for $\text{Br}(\mu\to e\gamma)$. In the latter case, Fig.~\ref{fig:lfv_max} (right), the MEG experiment sensitive to $\text{Br}(\mu\to e\gamma)\approx 10^{-13}$ will only be able to find a signal in case of a rather light SUSY spectrum in the region $m_0,m_{1/2} \lsim 200$~GeV. In the first case, Fig.~\ref{fig:lfv_max} (left), the part of the parameter space with $m_{1/2}\lesssim 200$~GeV is already excluded from $\text{Br}(\mu\to e\gamma)<1.2\times 10^{-11}$. The MEG experiment can probe a sizable part of the parameter space which partly overlaps with the region where one expects a significant number of LFV decays of $\tilde\chi^0_2$ at the LHC. Note, that a subspace of the parameter space which is probed by the LHC is complementary to the region probed by the MEG experiment. 

\section{Summary and Conclusions}\label{sec:summaryandconclusions}
In this paper, we have considered the lepton flavor violating decay branching ratios $\text{Br}(\mu\to e\gamma)$, $\text{Br}(\tau\to \mu\gamma)$, 
$\text{Br}(\tau\to e\gamma)$, as well as the LFV process $pp\to e \mu \tilde\chi^0_1 + X$ in mSUGRA for a broad class of lepton mass matrix textures that give nearly tribimaximal lepton mixing. The neutrino masses are normal hierarchical and become small due to the canonical type-I seesaw mechanism with
non-degenerate, mildly hierarchical right-handed neutrino masses. The textures exhibit large leptonic mixing that can originate from the charged lepton, the left- or the right-handed neutrino sector. We have studied the CP-conserving and the most general CP-violating case obtained by varying the CP-violating phases in the Lagrangian. In doing so, we focused on the SUSY benchmark scenario SPS1a'. We have determined the LFV decay rates for 72 qualitatively different sets of real texture patterns with a right-handed neutrino mass scale between $10^{11}\,\text{GeV}$ and $10^{15}\,\text{GeV}$. These 72 textures are taken from a previous systematic scan of real textures generating nearly tribimaximal mixing in a way similar to QLC. 

We have studied in detail the LFV rates for complex textures that are obtained by randomly varying the CP-violating phases in the Lagrangian in the most general way. The resulting complex textures exhibit nearly tribimaximal PMNS mixing angles and reproduce realistic lepton masses with a normal hierarchical neutrino mass spectrum. All complex
textures were, furthermore, subject to the requirement of having a perturbative Higgs sector. As expected for the SUSY seesaw mechanism, the electron electric dipole moment is several orders of magnitude below current bounds. We estimated and confirmed that promoting a real to a complex texture can lead to an enhancement of $\text{Br}(\mu\to e\gamma)$ of more than two orders of magnitude.

For the SUSY scenario SPS1a' and a heaviest right-handed neutrino mass of the order $2.5\times 10^{12}\,\text{GeV}$, the complex textures lead to rates $\text{Br}(\mu\to e\gamma)$ around the current experimental bound. The rare decay $\tau\to\mu\gamma$ could be measured in the future, whereas $\text{Br}(\tau\to e\gamma)$ seems to be out of reach of planned experiments. The predicted value of $\text{Br}(\tau\to\mu\gamma)$ is approximately independent of the chosen texture set and the CP phases. Potentially, it can therefore be used to determine the right-handed neutrino mass scale quite robustly if the SUSY mass spectrum is sufficiently well known. We have also compared the radiative LFV decay $\mu\to e \gamma$ with the event rate for the process $pp\to e\mu \tilde\chi^0_1 +X$ via LFV decays of the second lightest neutralino $\tilde \chi^0_2$ at the LHC. A sizable part of the mSUGRA parameter space that can be probed by this process at the LHC is found to be complementary to the one probed by the MEG experiment.

For the complex textures, we could not find a clear correlation between $\text{Br}(\mu\to e\gamma)$ and the PMNS mixing angles or phases of the low-energy neutrino sector such as the reactor angle $\theta_{13}$. The solar and atmospheric mixing angle as well as the two Majorana and the Dirac CP-violating phases are distributed over the whole allowed range without any clear preference for any angle or phase. Though present, only a few complex textures give extremely small values of the reactor angle $\theta_{13}\lesssim 1^\circ$.

We thus conclude that the inclusion of random CP-violating phases at the Lagrangian level typically erases possible correlations between the LFV BRs and the PMNS parameters such as $\theta_{13}$ that may occur for special points in parameter space. An interplay between a future measurement of the reactor angle and LFV BRs therefore seems highly model-dependent and would, even if the moduli of the Yukawa couplings are all fixed, require a detailed understanding of the possible CP-phases in the Lagrangian.

\section*{Acknowledgments}
The authors thank Werner Porod and Reinhold R\"uckl for very useful suggestions and help during the completion of this work and a careful reading of the manuscript. The authors thank Simon Albino for providing the code to calculate the LFV LHC process. G.S. is supported by the Federal Ministry of Education and Research (BMBF) under contract number 05HT6WWA. F.P. was supported in part by INFN under the program ``Fisica Astroparticellare'' and the Research Training Group 1147 ``{\it Theoretical Astrophysics and Particle Physics}'' of Deutsche Forschungsgemeinschaft.

\end{document}